\documentclass{article}
\usepackage{ijcai25}

\usepackage{times}
\usepackage{soul}
\usepackage{url}
\usepackage[hidelinks]{hyperref}
\usepackage[utf8]{inputenc}
\usepackage[small]{caption}
\usepackage{graphicx}
\usepackage{amsmath}
\usepackage{amsthm}
\usepackage{booktabs}
\usepackage{algorithm}
\usepackage{algorithmic}
\usepackage[switch]{lineno}

\usepackage{amsfonts} %
\usepackage{tabularx, dcolumn} %
\usepackage[dvipsnames,table]{xcolor}
\usepackage{tikz}
\usepackage{worldflags}
\usepackage{subcaption}
\RequirePackage{etex} %
\hypersetup{bookmarksdepth=subsection} %

\usepackage{diagbox}

\newcommand{\pill}[3][yellow]{%
  \tikz[baseline=(X.base)] \node[draw=none, fill=#1, rounded corners=5pt, text=black, inner sep=0pt, minimum height=1.2em, minimum width=4em, align=center] (X) {#2};%
}
\newcommand{\icon}[1]{\makebox[0.4cm]{#1}}
\definecolor{candidate}{named}{NavyBlue}  %
\definecolor{platform}{named}{OliveGreen}   %
\definecolor{question}{named}{Plum} %
\definecolor{high}{RGB}{139, 69, 19}
\definecolor{medium}{HTML}{ffb442}
\definecolor{high}{HTML}{f06542}

\usepackage[most]{tcolorbox}
\usepackage{booktabs}
\usepackage{fontawesome5}
\usepackage{makecell}
\usepackage{natbib}

\definecolor{mylightgrey}{HTML}{d3d3d3} %
\definecolor{mygreen}{HTML}{8fb935} %
\definecolor{myyellow}{HTML}{e6e22e} %
\definecolor{myorange}{HTML}{e09c3b} %
\definecolor{myred}{HTML}{e64747} %

\definecolor{mygreen1}{HTML}{b8db85} %
\definecolor{mygreen2}{HTML}{a5d36b} 
\definecolor{mygreen3}{HTML}{8fb935} 

\definecolor{myyellow1}{HTML}{f1eb9c} %
\definecolor{myyellow2}{HTML}{ebdb5b}
\definecolor{myyellow3}{HTML}{e6e22e} 

\definecolor{myred1}{HTML}{f3b5b5} %
\definecolor{myred2}{HTML}{ea8b8b}
\definecolor{myred3}{HTML}{e64747} 

\definecolor{Blues1}{HTML}{f7fbff}  %
\definecolor{Blues2}{HTML}{deebf7}
\definecolor{Blues3}{HTML}{c6dbef}
\definecolor{Blues4}{HTML}{9ecae1}
\definecolor{Blues5}{HTML}{6baed6}
\definecolor{Blues6}{HTML}{4292c6}
\definecolor{Blues7}{HTML}{2171b5}
\definecolor{Blues8}{HTML}{08519c}
\definecolor{Blues9}{HTML}{08306b}  %

\newcommand{\riskcell}[1]{
    \ifthenelse{\equal{#1}{Low}}{\cellcolor{myyellow!80}}{}
    \ifthenelse{\equal{#1}{Medium}}{\cellcolor{myorange!80}}{}
    \ifthenelse{\equal{#1}{High}}{\cellcolor{myred!80}}{}
    #1
}

\newcommand{\colorrisk}[1]{
    \ifnum#1=1 \cellcolor{myred} %
    \else\ifnum#1=2 \cellcolor{myorange} %
    \else\ifnum#1=3 \cellcolor{mygreen} %
    \else\ifnum#1=4 \cellcolor{mylightgrey} %
    \fi\fi\fi\fi
}

\newcolumntype{R}[2]{%
    >{\adjustbox{angle=#1,lap=\width-(#2)}\bgroup}%
    l%
    <{\egroup}%
}

\newcolumntype{L}[1]{>{\raggedright\arraybackslash}p{#1}}

\renewcommand{\paragraph}[1]{\vspace{0.5em}\noindent\textbf{#1}} %

\newcommand{\posColor}[1]{\cellcolor{mygreen!#1}}
\newcommand{\negColor}[1]{\cellcolor{myred!#1}}

\title{Recommender Systems for Democracy: Toward Adversarial Robustness in Voting Advice Applications}

\author{
Frédéric Berdoz$^1$
\and
Dustin Brunner\and
Yann Vonlanthen$^1$\And
Roger Wattenhofer\\
\affiliations
ETH Zurich, Switzerland\\
\emails
fberdoz@ethz.ch,
yvonlanthen@ethz.ch,
wattenhofer@ethz.ch
}

\begin{document}

\maketitle

\footnotetext[1]{Corresponding authors.}
\setcounter{footnote}{1}

\begin{abstract}

Voting advice applications (VAAs) help millions of voters understand which political parties or candidates best align with their views. This paper explores the potential risks these applications pose to the democratic process when targeted by adversarial entities. In particular, we expose 11 manipulation strategies and measure their impact using data from Switzerland’s primary VAA, Smartvote, collected during the last two national elections. We find that altering application parameters, such as the matching method, can shift a party’s recommendation frequency by up to 105\%. Cherry-picking questionnaire items can increase party recommendation frequency by over 261\%, while subtle changes to parties’ or candidates’ responses can lead to a 248\% increase. To address these vulnerabilities, we propose adversarial robustness properties VAAs should satisfy, introduce empirical metrics for assessing the resilience of various matching methods, and suggest possible avenues for research toward mitigating the effect of manipulation. Our framework is key to ensuring secure and reliable AI-based VAAs poised to emerge in the near future.
\end{abstract}

\section{Introduction}

Recent advances in information technology have significantly transformed our daily lives. One area that remains relatively underexplored is digital democracy, which integrates digital innovations into the political system. Among the most notable developments in this field is the emergence of Voting Advice Applications (VAAs). VAAs provide voters with personalized recommendations on which parties or candidates best align with their preferences and policy stances. 
VAAs exist in as many as 30 countries across the world, including the USA, Canada, Australia, as well as many European countries~\citep{teran2020voting}.  
Interestingly, the legal basis of VAAs varies widely from country to country, ranging from publicly governed and regulated entities to loosely controlled private associations~\citep{garzia2012under-review}.  Strikingly, almost every country chooses a different method to match voters to candidates~\citep{louwerse2014designeffects}. 
In countries where VAAs are currently in use, they are often consulted by 10-50\% of voters~\citep{teran2020voting}, making them a highly popular source of information. On top of that, the advice provided by these applications has been shown to significantly influence both voter turnout and voter decisions~\citep{munzert2021meta}. In Switzerland specifically, \citet{germann2019getting} showed that the VAA mobilized 58,000 additional voters in 2007, while \citet{ladner2010voting} reported that 67\% of the users had stated that the VAA had influenced their voting behavior. 
The profound impact of VAAs has gone as far as triggering a shift from representative to promissory democracy, in which VAA profiles are interpreted as electoral promises~\citep{ladner2016promissory}. This transition occurred without requiring any changes to constitutional or legal frameworks.
While the benefits of VAAs are undeniable and well-documented~\citep{munzert2021meta}, for the first time, this study aims to shed light on their potential vulnerabilities. Specifically, we seek to quantify the impact that a hypothetical adversarial actor could have on the recommendations. 
Toward this goal, we focus our analysis on Smartvote, Switzerland's primary VAA. In 2023, Smartvote was used by up to 20\% of eligible Swiss voters, up from 17\% in 2011. In 2023, a total of 2.1 million voting advice reports were created~\citep{smartvote2024website}. Our contributions:
\begin{enumerate}
    \item We propose three adversarial robustness properties for VAAs. Namely, robustness against manipulation by (i) candidates and parties, (ii) platform operators, and (iii) question designers (Section~\ref{sec:background}).
    \item We empirically demonstrate the importance of these robustness properties by leveraging two comprehensive datasets collected by Smartvote during the Swiss national elections of 2019 and 2023. We uncover a total of 11 vulnerabilities through which adversaries could manipulate the recommendations (Table~\ref{tab:dangers_overview} and App.~\ref{app:vulnerabilities}).
    \item Based on the highest-risk vulnerabilities, we suggest 9 metrics to compare the adversarial robustness of existing and newly proposed matching methods (Section~\ref{sec:metrics}). 
    \item  Finally, with input from Politools, the non-profit organization behind Smartvote, we propose research directions to mitigate these vulnerabilities, enabling the development of more robust VAAs in the future (Section~\ref{sec:mitigations}).
\end{enumerate}
\begin{table*}[t]
    \centering
    \renewcommand{\arraystretch}{1.2} %
        \small
        \begin{tabularx}{\textwidth}{
        p{2.8cm} 
        p{2.5cm} 
        >{\centering\arraybackslash}p{1.1cm}
        >{\centering\arraybackslash}X
        >{\centering\arraybackslash}X
        >{\centering\arraybackslash}p{2cm}
        >{\centering\arraybackslash}X
        >{\centering\arraybackslash}X
        >{\centering\arraybackslash}X
        }
        \toprule
         Vulnerability & Adversary Type  & Code & Section  & Data &  Benefactors & Visibility Gain & Likelihood & Impact  \\
        \midrule
         Answer Optimization      & Candidates             & \textcolor{candidate}{AO}   & \ref{sec:vulnerabilities:candidates}    & \faCheck & \icon{\faUser} \icon{} \icon{} \icon{}                   & 259\% & \pill[GreenYellow]{\emph{Low}}     & \pill[high]{\emph{High}} \mbox{}             \\
         Answer Calibration       & Candidates             & \textcolor{candidate}{AC}   & \ref{sec:vulnerabilities:candidates}    & \faTimes & \icon{\textcolor{lightgray}{\faUser}} 
                                                                                                                                                \icon{\textcolor{lightgray}{\faList}} \icon{\faFlag} 
                                                                                                                                                \icon{\textcolor{lightgray}{\faBalanceScale}}            & 248\% & \pill[high]{\emph{High}}            & \pill[high]{\emph{High}} \mbox{}             \\
         Diversification          & Candidates (Party)     & \textcolor{candidate}{DIV}  & \ref{sec:vulnerabilities:candidates}    & \faTimes & \icon{} \icon{} \icon{\faFlag} \icon{}                   & 345\% & \pill[medium]{\emph{Medium}}            & \pill[high]{\emph{High}}\mbox{}      \\
         List Centralization      & Candidates (Party)     & \textcolor{candidate}{LC}   & App. \ref{app:vulnerabilities}          & \faTimes & \icon{} \icon{\faList} \icon{} \icon{}                   & - & \pill[GreenYellow]{\emph{Low}}     & \pill[GreenYellow]{\emph{Low}}\mbox{}       \\
         Matching Method          & Platform operator      & \textcolor{platform}{MM}    & \ref{sec:vulnerabilities:platform}      & \faTimes & \icon{} \icon{} \icon{}\icon{\faBalanceScale}            & 105\% & \pill[medium]{\emph{Medium}}    & \pill[medium]{\emph{Medium}}\mbox{}              \\
         Question Ordering        & Platform operator      & \textcolor{platform}{QO}    & App. \ref{app:vulnerabilities}      & \faCheck & \icon{} \icon{}\icon{\faFlag} \icon{}                    & \phantom{00}6\% & \pill[GreenYellow]{\emph{Low}}            & \pill[GreenYellow]{\emph{Low}}\mbox{}       \\
         Weight Selection         & Platform operator      & \textcolor{platform}{WS}    & \ref{sec:vulnerabilities:platform}      & \faTimes & \icon{} \icon{} \icon{}\icon{\faBalanceScale}            & {\footnotesize$\approx$}15\% & \pill[GreenYellow]{\emph{Low}}    & \pill[GreenYellow]{\emph{Low}} \mbox{}     \\
         Similarity Score         & Platform operator      & \textcolor{platform}{SS}    & \ref{sec:vulnerabilities:platform}      & \faTimes & \icon{} \icon{} \icon{\faFlag} 
                                                                                                                                                \icon{\textcolor{lightgray}{\faBalanceScale}}            & - & \pill[medium]{\emph{Medium}}    & \pill[GreenYellow]{\emph{Low}}\mbox{}       \\
         Tie-breaking             & Platform operator      & \textcolor{platform}{TB}    & App. \ref{app:vulnerabilities}      & \faTimes & \icon{\faUser} \icon{} \icon{} \icon{}                   & 210\% & \pill[medium]{\emph{Medium}}    & \pill[medium]{\emph{Medium}}\mbox{}      \\
         Question Favoritism      & Question designer      & \textcolor{question}{QF}    & \ref{sec:vulnerabilities:questions}     & \faCheck & \icon{} \icon{} \icon{\faFlag} \icon{}                   & 261\% & \pill[GreenYellow]{\emph{Low}}     & \pill[high]{\emph{High}}\mbox{}              \\
         Question Correlation     & Question designer      & \textcolor{question}{QC}    & \ref{sec:vulnerabilities:questions}     & \faTimes & \icon{} \icon{} \icon{\faFlag} \icon{}                   & - & \pill[medium]{\emph{Medium}}     & \pill[medium]{\emph{Medium}}\mbox{}      \\
        \bottomrule
        \end{tabularx}
        \caption{Overview of the main vulnerabilities associated with each type of adversary, with type-specific color codes for reference in the paper. The \emph{Data} column indicates whether a strategy exploiting that vulnerability requires knowledge of the voters' or candidates' answers. For \emph{Benefactors}, \icon{\faUser} denotes single candidates, \icon{\faList} denotes lists, \icon{\faFlag} denotes parties, and \icon{\faBalanceScale} denotes party coalitions (i.e., left, center, right as shown in Figure~\ref{fig:weight_selection}). The primary benefactor is highlighted in \textbf{black}, secondary benefactors are shown in \textcolor{lightgray}{\textbf{gray}}. The \emph{Visibility Gain} factor indicates its best-case potential relative increase in visibility in the VAA if that vulnerability is exploited, as estimated by our experiments throughout the paper (left blank if no experiment was conducted). The table also includes a subjective assessment of the \emph{Likelihood} and \emph{Impact} of each strategy. %
        }
    \label{tab:dangers_overview}
\end{table*}

\section{Background}
\label{sec:background}
Most popular VAAs use a set of questions $\smash{Q = \{q_t\}_{t=1}^{N_q}}$ to position both candidates $\smash{C = \{c_j\}_{j=1}^{N_c}}$ and voters $\smash{V = \{v_i\}_{i=1}^{N_v}}$ within the high-dimensional Euclidean space $ \mathbb{R}^{N_q} $. Formally, each question $ q_t: {V \cup C \to A_t}$ assigns an answer to a given voter or candidate, with $A_t \subseteq \mathbb{R}$ being the set of allowable answers for that question (generally discrete and bounded). 
For example, the question “Are VAAs robust?” might map the answers “No”, “Rather no”, “Rather yes”, and “Yes” to the numerical values $0$, $25$, $75$, and $100$, respectively. Additionally, for each question $q_t$, voters can typically choose a numerical weight within a set of allowable values $ W_t \subseteq \mathbb{R} $ to reflect how important each question is to them. This weight is formally represented as a mapping $w_t: {V \to W_t}$.
Given a voter-candidate pair $(v_i, c_j)$ and their respective answer and weight vectors ${\mathbf{v}_i = [q_1(v_i), ..., q_{N_q}(v_i)]^T}$, ${\mathbf{c}_j=[q_1(c_j),...,q_{N_q}(c_j)]^T}$ and ${\mathbf{w}_i = [w_1(v_i), ..., w_{N_q}(v_i)]^T}$, the VAA computes a similarity score $s(v_i, c_j)$ between $v_i$'s and $c_j$'s opinions using a predefined weighted distance function $d(\mathbf{v}_i, \mathbf{w}_i, \mathbf{c}_j)$, with ${d: \mathbb{R}^{N_q} \times \mathbb{R}^{N_q}  \times \mathbb{R}^{N_q} \to \mathbb{R}_+}$. Lastly, for each voter $v_i$, the VAA provides a ranking $\mathbf{r}_i\in\mathcal{R}(C)$ based on these similarity scores, with $\mathcal{R}(C)$ the set of total orders on $C$. See Table \ref{tab:vaa_database_overview} in the Appendix for a summary of how the most popular VAAs align with this framework. As some of our analysis will concern parties and lists, we also account for the fact that candidates can belong to exactly one party $p \in P$ and one list $l\in L$, with $P$ and $L$ being the set of all parties and lists, respectively. In Swiss National Council elections, lists are party- or coalition-specific slates of candidates from which voters choose or modify their preferred selections (see Appendix~\ref{app:smartvote:swisselections} for more details).
A canonical set of properties that any safe VAA must satisfy commonly includes \citep{garziamarschall2014matching}:
\begin{enumerate}
    \item[(R)] \emph{Reproducibility}: The VAA produces reproducible recommendations, enabling users to verify the system's reliability.
    \item[(I)] \emph{Interpretability}: The rationale behind the VAA's recommendations is easily understandable and intuitive to users, including those with less technical expertise.
    \item[(T)] \emph{Transparency}: The VAA's matching algorithm and all factors influencing recommendations are open-source.
    \item[(F)] \emph{Fairness}: The VAA is purely issue-based and does not consider any other characteristics of voters or candidates.
    \item[(E)] \emph{Explainability}: Voters receive clear and intuitive explanations for candidate or list recommendations.
    \item[(P)] \emph{Privacy}: The VAA ensures the privacy and anonymity of users' responses and preferences.
\end{enumerate}
Although the importance of these properties is clear, they do not offer protection against malicious actors (i.e., adversaries) aiming to manipulate the recommendations to favor a particular candidate or party. From the above definitions, one can identify three potential types of such adversaries: (i) The \textcolor{candidate}{\textbf{candidates}} providing their answer vectors, (ii) the \textcolor{platform}{\textbf{platform operator}} in charge of choosing $d$, $\smash{\{A_t\}_{t=1}^{N_q}}$, $\smash{\{W_t\}_{t=1}^{N_q}}$ and all other aspects related to VAA's interface  (such as question ordering, tie-breaking, etc.), and (iii) the \textcolor{question}{\textbf{question designers}} writing the questions $Q$.\footnote{For Smartvote, the non-profit association Politools is responsible for selecting the questions and operating the platform~\citep{smartvote2024website}.}
In Section \ref{sec:vulnerabilities}, we analyze the primary dangers associated with each type of adversary, grounding our analysis in the two datasets from Smartvote presented in Section \ref{sec:dataset}. 
Then, in Section \ref{sec:mitigations}, we propose solutions to mitigate these risks. 

\section{Dataset}
\label{sec:dataset}
We empirically evaluate our claims using two comprehensive datasets collected by Smartvote~\citep{smartvote2024website}, which include questionnaire responses and metadata from both voters and candidates in the 2019 and 2023 Swiss National Council elections. In both elections, approximately 85\% of electable candidates participated by completing the questionnaire, and around 20\% of eligible Swiss voters used Smartvote for voting recommendations. These recent datasets provide a solid foundation for analyzing VAA robustness, capturing a significant portion of both voters and candidates.
Smartvote contains $N_q=75$ questions with ${A_t = \{0,25, 75, 100\}}$ for questions ${1\leq t \leq 60}$ (policy questions), ${A_t = \{0,17,33,50,67,83,100\}}$ for ${61\leq t \leq 67}$ (value questions) and ${A_t = \{0,25,50, 75, 100\}}$ for ${68\leq t \leq 75}$ (budget questions). For all questions, the allowable values for the weights are $W_t= \{0, 0.5, 1, 2\}$, with $1$ being the default value for answered questions and $0$ the value automatically assigned to any unanswered question. The distance metric used in Smartvote is the L2 distance 
\begin{equation}
\label{eq:L2_distance}
{d_{\mbox{\scriptsize L2}}(\mathbf{v}_i, \mathbf{w}_i, \mathbf{c}_j)=\sqrt{\sum_{t=1}^{N_q} (\mathbf{w}_{i,t}(\mathbf{v}_{i,t} - \mathbf{c}_{j,t}))^2}},
\end{equation}
which is used to compute the normalized similarity scores
\begin{equation}
\label{eq:similarity_score}
s(v_i, c_j)=100\cdot\left(1-\frac{d_{\mbox{\scriptsize L2}}(\mathbf{v}_i, \mathbf{w}_i, \mathbf{c}_j)}{d_{\mbox{\scriptsize L2}}(100 \cdot \mathbf{1}_{N_q}, \mathbf{w}_i, \mathbf{0}_{N_q})}\right),
\end{equation}
where $\mathbf{1}_{N_q}$ (respectively $\mathbf{0}_{N_q}$) denote the one-valued (respectively zero-valued) $N_q$ dimensional vector. In addition to candidate rankings, Smartvote also provides a list ranking by averaging the similarity scores of all candidates on each list $l\in L$, i.e., $\smash{s(v_i, l) = \frac{1}{|l|}\sum_{c\in l}s(v_i,c)}$. For a more detailed description of the Swiss political system and Smartvote, we refer the reader to Appendix~\ref{app:smartvote}. In Appendix~\ref{app:dataset}, we provide a comprehensive description of the preprocessing applied to the two datasets, as well as an exploratory data analysis.
We conducted all analyses and experiments on both datasets, but present results from the more recent 2023 dataset, as the overall findings are consistent across both elections.

\section{Vulnerabilities}
\label{sec:vulnerabilities}

While Smartvote satisfies in large part\footnote{The \emph{fairness} and \emph{reproducibility} properties of Smartvote are not fully met, as they break ties using last names and allowed some candidates to overwrite their initial answers on a few questions.} all the safety properties listed in Section~\ref{sec:background}, its robustness to adversarial entities remains unclear.
In this section, we analyze the key strategies that the different types of adversaries might use to increase the visibility of a particular candidate or party. Given a set of candidates $C$ and a set of recommendations (i.e., rankings) ${R_C = \{\mathbf{r}_i \in \mathcal{R}(C) \mid v_i \in V\}}$, we define the $k\text{-\textbf{visibility of a candidate}}$ $\nu_k(c \mid C)$ as the frequency with which candidate $c$ appears in the top $k$ positions of the rankings $R_C$. Additionally, we define the $k\text{-\textbf{visibility of a party}}$ $\nu_k(p \mid P)$ as the fraction of the top $k$ recommendations that are occupied by members of that party. Finally, we define the $k\text{-\textbf{visibility of a list}}$ $\nu_k(l \mid L)$ as the frequency with which $l$ appears in the top $k$ positions of the list rankings ${R_L= \{\mathbf{r}_i \in \mathcal{R}(L) \mid v_i \in V\}}$. Throughout this work, unless specified otherwise, we set $k$ to the number of seats allocated to the candidate's state\footnote{Usually referred to as a \emph{canton} in Switzerland} in the National Council, for both candidate and party visibility. For lists, we use $k=1$ by default, as voters can only vote for one list. These default values also correspond to the number of candidates and lists visually put forward by Smartvote. Due to their specificity, we discuss the list centralization (\textcolor{candidate}{LC}), the question ordering (\textcolor{platform}{QO}), and the tie-breaking (\textcolor{platform}{TB}) vulnerabilities in Appendix~{\ref{app:vulnerabilities}}. 

\subsection{Candidates and Parties}
\label{sec:vulnerabilities:candidates}

\subsubsection{Answer Optimization (\textcolor{candidate}{AO})}
\begin{figure}[t!]
    \centering
    \includegraphics[width=1\linewidth]{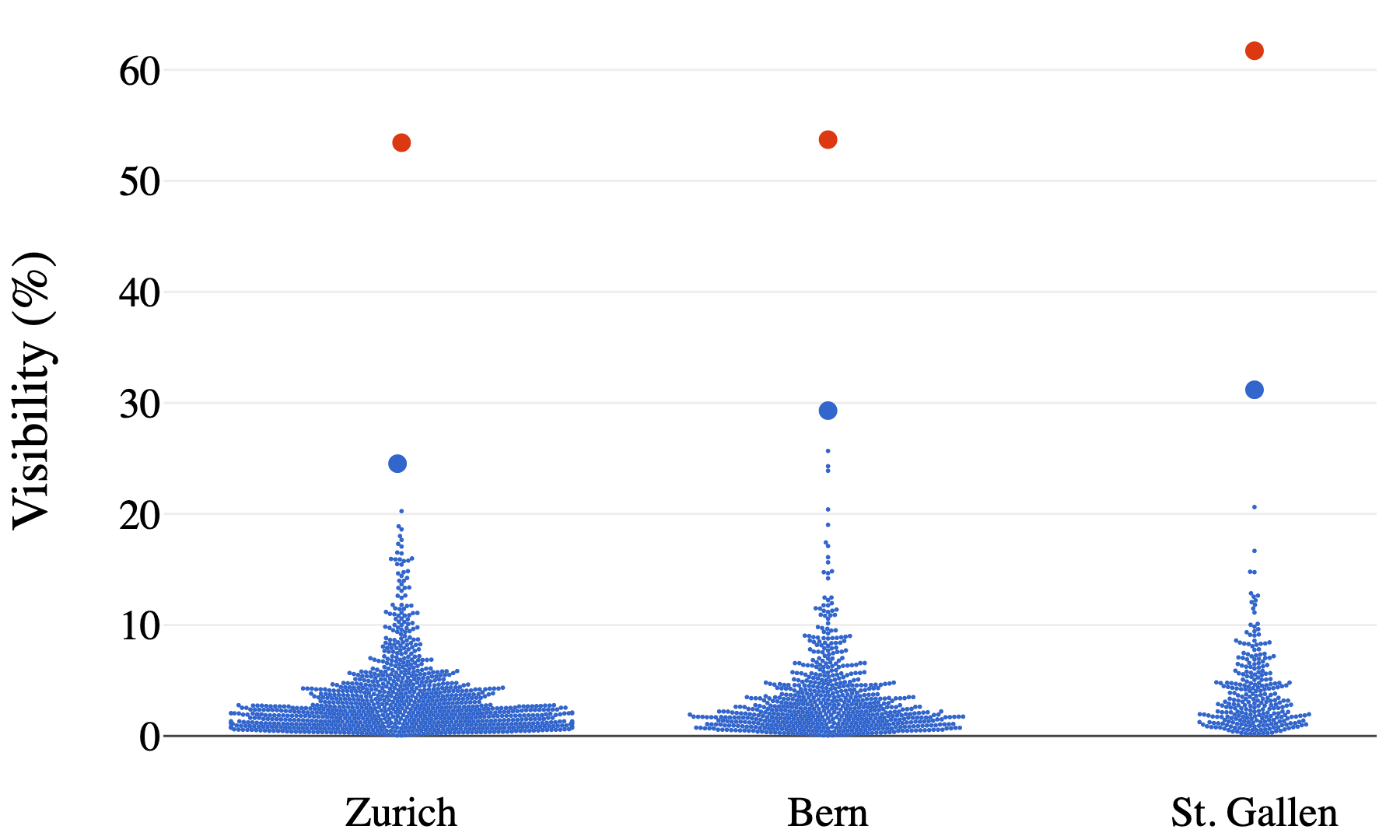}
    \caption{Visibility of crafted candidates (red) compared to all other candidates (blue) in the states of Zurich ($k=36$), Bern ($k=24$), and St. Gallen ($k=12$). The larger dots highlight the crafted and actual most visible candidates.}
    \label{fig:optimization}
\end{figure}
We start by investigating the potential for a single candidate to manipulate their answers to increase their popularity. 
The computation of the provably optimal candidate is of combinatorial complexity and thus infeasible, as pointed out by \citet{etter2014mining}. However, we can find an approximate solution through randomized optimization. For each state, we craft an artificial candidate $c^*$ using \textit{simulated annealing}~\citep{kirkpatrick1983optimization} and optimizing $\nu_k(c^*\mid {C \cup \{c^*\}})$. 
In almost all states, the crafted candidate appears in more than 50\% of top $k$ recommendations, significantly outperforming the previously crafted candidate by \citet{etter2014mining}, as well as any actual candidate.\footnote{Note that \citet{etter2014mining} set $k = 50$ in the popularity metric, while for us $k \in \{1, \cdots, 36\}$. Our result is thus strictly stronger.} 
Figure~\ref{fig:optimization} shows that the crafted candidates in the states of Zurich, Bern, and St. Gallen easily surpass their competition in terms of visibility. Table~\ref{tab:optimization_overview} in Appendix~\ref{app:vulnerabilities} contains the popularity of our best crafted candidate for each state, as well as a comparison with other optimization strategies. Specifically, it demonstrates that the visibility of candidates crafted using only 1\% of the voters' data is nearly as high as those optimized with the full dataset, achieving 51.70\%, 50.66\%, and 52.55\% in Zurich, Bern, and St. Gallen, respectively. 
The analysis of the crafted candidate's profile reveals that almost no questions are answered on the answer spectrum's extremities (e.g., only 2 out of 75 answers for the crafted candidate in Zurich). This points to a systematic bias toward candidates with moderate positions. We investigate this lead next.

\subsubsection{Answer Calibration (\textcolor{candidate}{AC})}
\begin{figure}[t]
    \centering
    \includegraphics[width=1\linewidth]{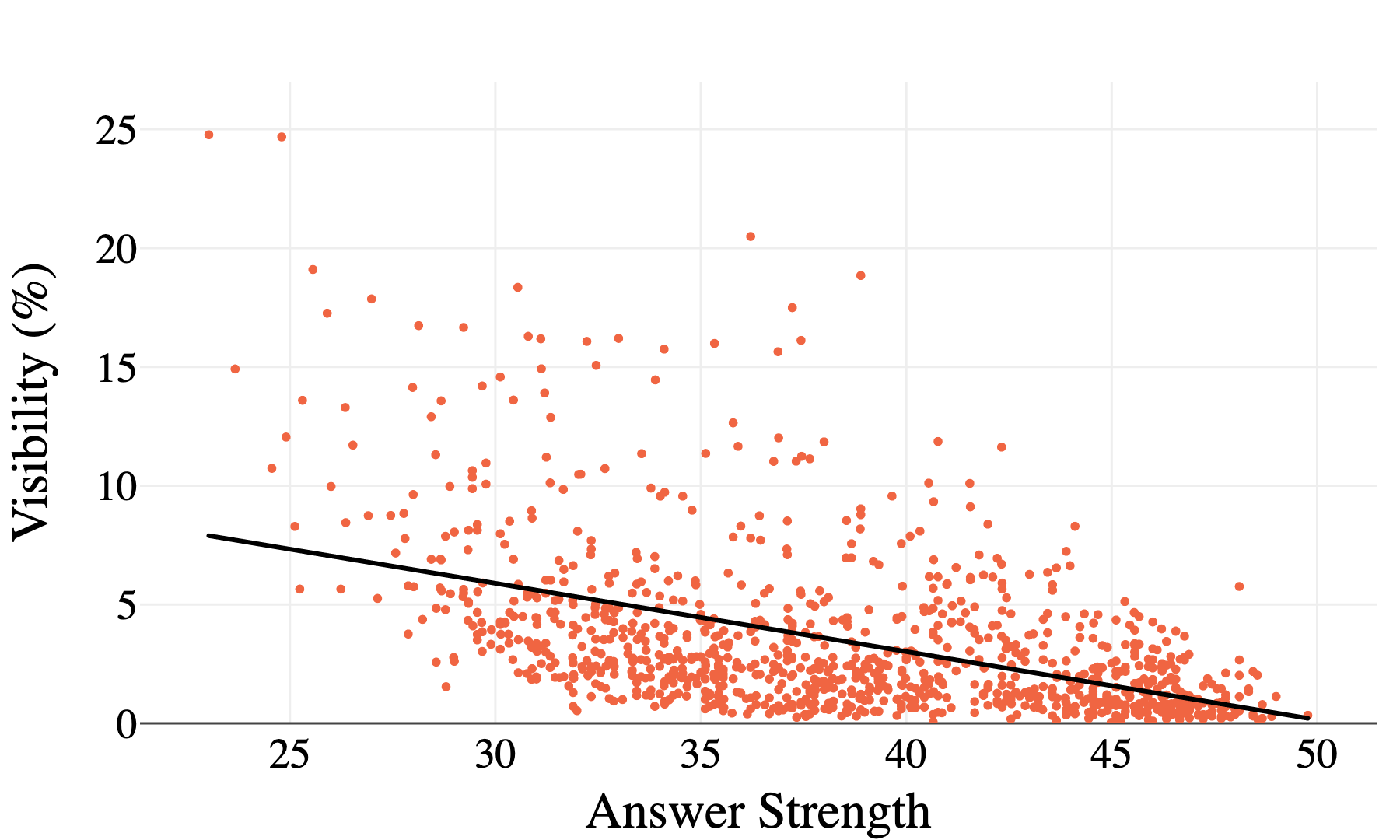}
    \caption{Relationship between the answer strength of candidates, as defined in Eq.~\eqref{eq:answer_strength}, and their visibility in the state of Zurich ($k=36$). Each dot shows a candidate and the black line represents an ordinary least squares trend line.}
    \label{fig:answer_strength_recommendations}
\end{figure}

\begin{figure}[t]
    \centering
    \includegraphics[width=1\linewidth]{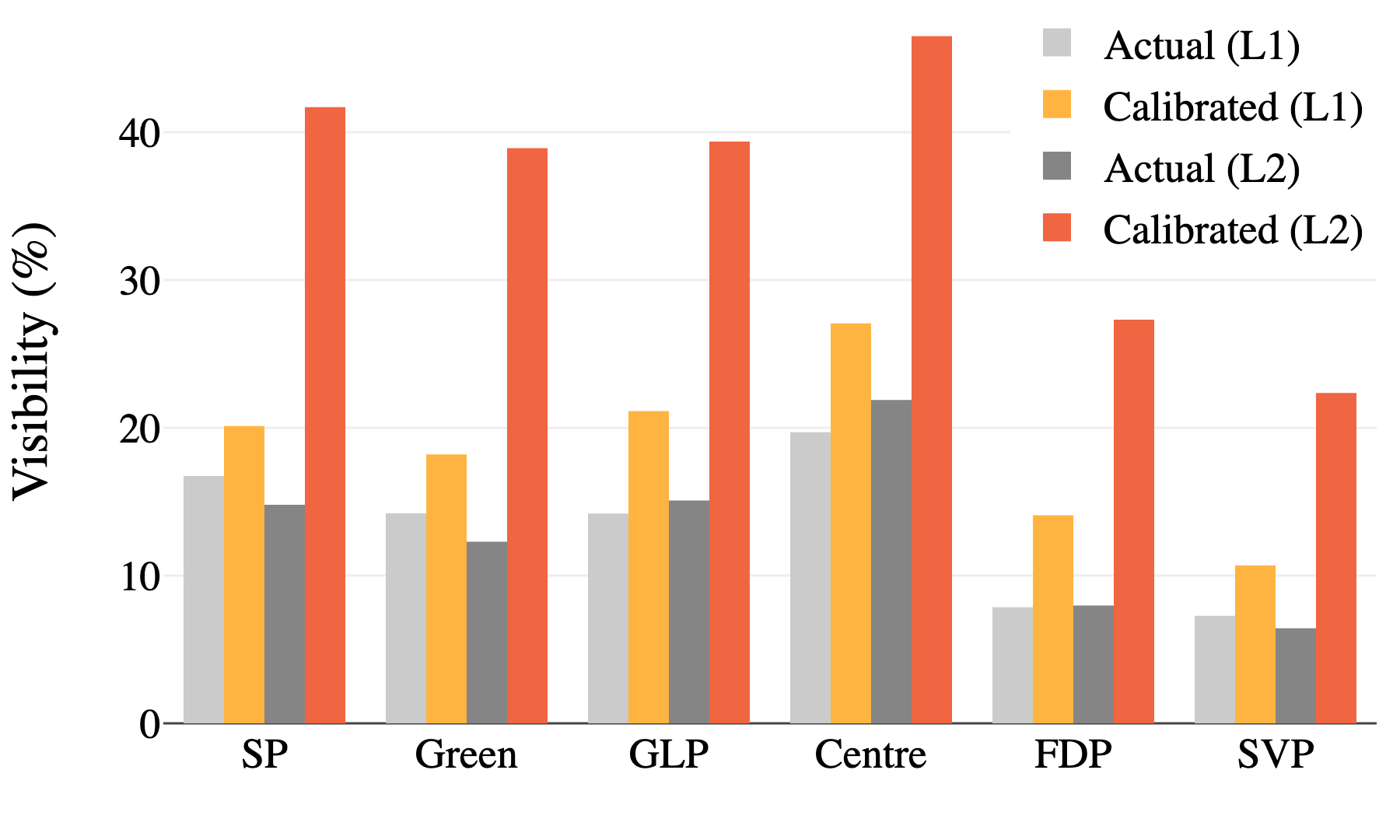}
    \caption{Comparison of actual and calibrated party visibility using the L1 and L2 distance metrics. To simulate this scenario, the answer profiles of all candidates in the party were adjusted to weaken their responses (e.g., changing all “Yes” to “Rather yes”), and the recommendations were recalculated using the L1 and L2 distance metrics.}
    \label{fig:neutral_answer_strategy_recommendation_shares}
\end{figure}

In Smartvote, candidates are provided with four or more response options. They can deliberately choose to respond “strongly” by selecting answers at the poles ($0$ or $100$) or “moderately” by choosing options closer to the middle of the answer spectrum ($25$ or $75$).\footnote{Answering moderately can be used to indicate a nuanced position, openness to compromise, or ambivalence. As such, the added expressivity is regarded to be beneficial~\citep{batterton2017likert}.} 
We define the \textbf{strength} $\sigma$ of an answer $\mathbf{c}_j$ by its deviation from the neutral position in absolute value, i.e., \begin{equation}
\label{eq:answer_strength}
\sigma(\mathbf{c}_j)= \frac{1}{N_q}\sum_{t=1}^{N_q}| \mathbf{c}_{j,t} - \frac{1}{2}(\max A_t + \min A_t)|.
\end{equation} 
In Figure~\ref{fig:answer_strength_recommendations}, we find that in Smartvote, candidates with moderate answers (i.e., lower answer strength) are recommended significantly more often. This concerning trend suggests that candidates can artificially boost their visibility by providing moderate answers to all questions. This strategy is particularly problematic because it can be executed with minimal deviation from the true candidate's position, making it difficult to detect. Figure~\ref{fig:neutral_answer_strategy_recommendation_shares} reveals that with the current distance metric used in Smartvote ($d_{\mbox{\scriptsize L2}}$), some parties can increase their visibility fourfold by unilaterally adopting this strategy.

\subsubsection{Diversification (\textcolor{candidate}{DIV})}
\begin{figure}[t]
    \centering
    \includegraphics[width=1\linewidth]{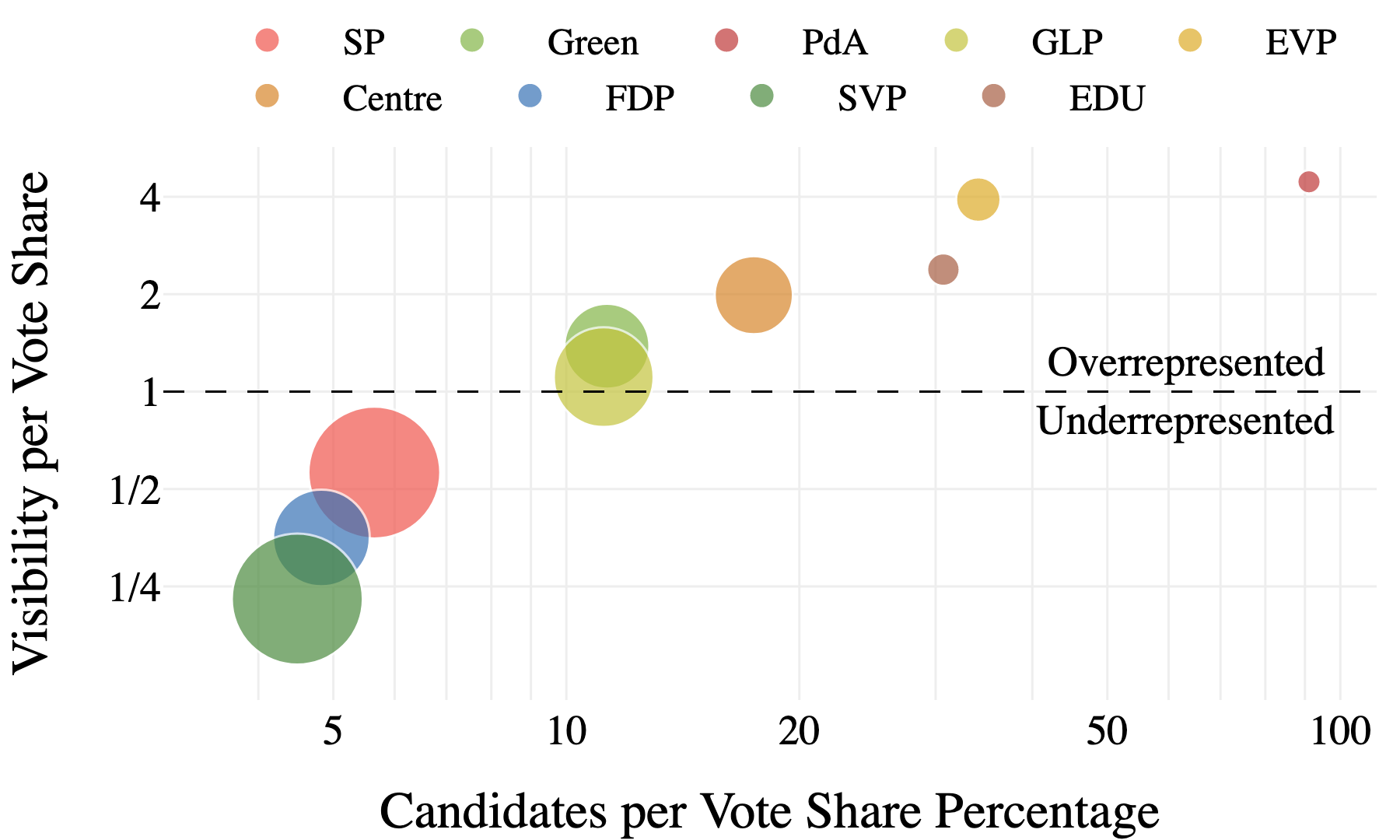}
    \caption{Relationship between the number of candidates per percent of vote share and the ratio of visibility to vote share for parties in the state of Zurich. The size of each dot represents the vote share of the corresponding party. Vote shares are calculated based on the votes received by candidates participating in Smartvote for the 2023 National Council election. Exact values can be found in the column \emph{Vote Share (adjusted)} of Table~\ref{tab:parties} in the Appendix.}
    \label{fig:candidate_diversification_correlation}
\end{figure}

Figure~\ref{fig:candidate_diversification_correlation} shows that parties with more candidates relative to their vote share tend to receive disproportionately more recommendations on Smartvote. This significant correlation suggests that having more candidates can skew recommendations, thereby providing an artificial advantage in voter outreach and potentially electoral success.

\subsection{Platform Designers}
\label{sec:vulnerabilities:platform}

\subsubsection{Matching Method (\textcolor{platform}{MM})}
\citet{louwerse2014designeffects} show how sensitive recommendations are to changes in the matching method. We extend these findings by quantitatively evaluating the bias and accuracy of each distance function in Table~\ref{tab:method_comparison}. Additionally, in Table~\ref{tab:method_party_visibilities} of the Appendix, we show that some methods can disproportionately favor candidates at either end of the political spectrum.

\subsubsection{Weight Selection (\textcolor{platform}{WS})}
\begin{figure}[t]
   \centering
   \includegraphics[width=1\linewidth]{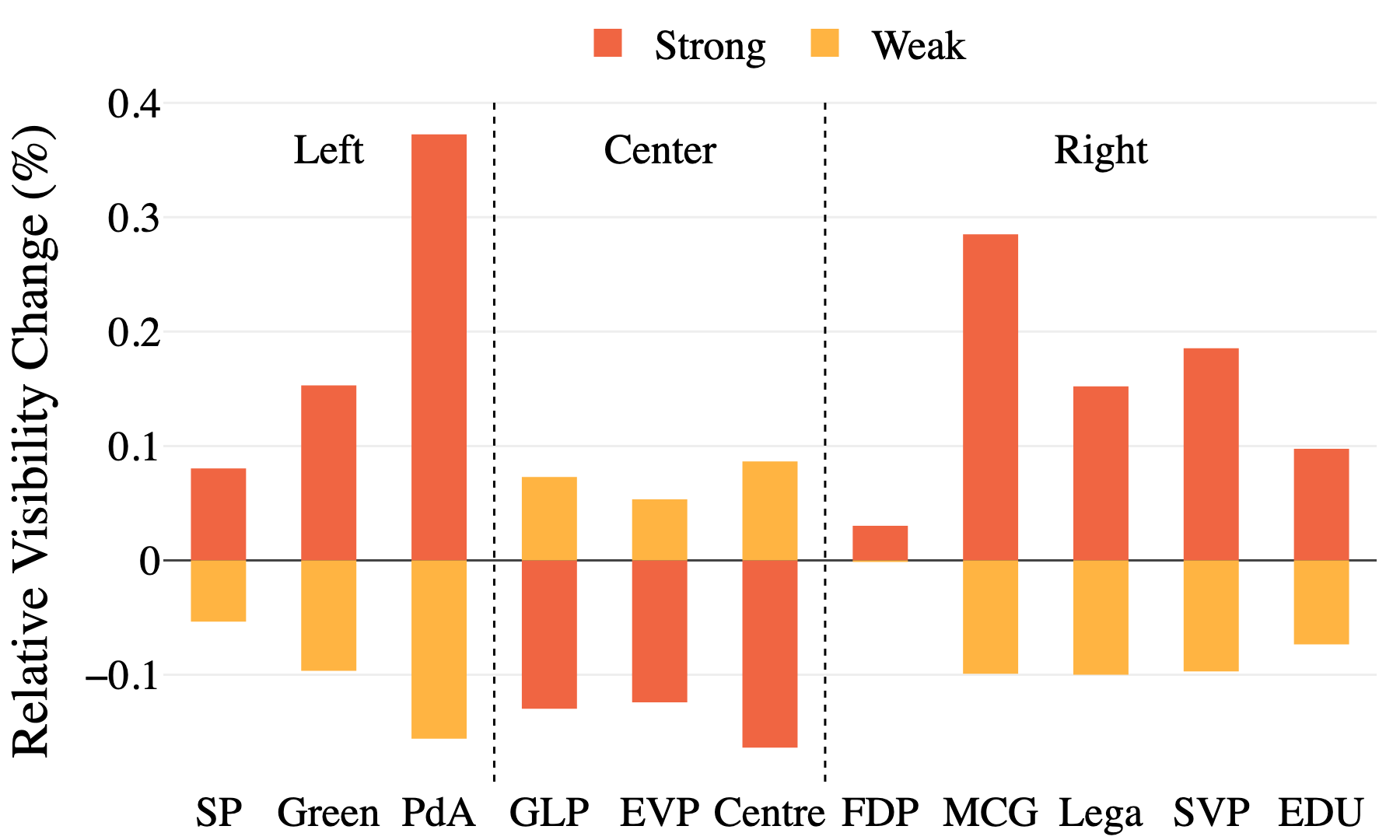}
   \caption{Relative visibility change of all parties if the available question weights are set to $\smash{W_t=\{0,\frac{1}{10}, 1, 10\}}$ (strong) or $\smash{W_t=\{0, \frac{9}{10},1, \frac{10}{9}\}}$ (weak). The visibility of each party is computed using only the voters that have weighted at least one question. Parties are listed according to their parliamentary seating arrangement, with traditional larger coalitions (left, center, right) shown at the top. As observed, the actual numerical value of the weights can significantly favor certain coalitions, with center parties benefiting from weak weights and left- and right-wing parties from strong weights.}
   \label{fig:weight_selection}
\end{figure}

In Smartvote, voters have the option to decrease or increase the weight of each question $q_t$, but without knowing the actual numerical weights $\smash{W_t=\{0, \frac{1}{2}, 1, 2\}}$ corresponding to these actions.\footnote{These values are available on the \emph{About} page on Smartvote, but they are not displayed directly alongside the questions.} Figure~\ref{fig:weight_selection} displays the relative change of the main parties' visibility (among voters that have weighted at least one question) if these values are changed. 

\subsubsection{Similarity Score (\textcolor{platform}{SS})}
Apart from determining the ranking $\mathbf{r}_i$, the similarity scores $s(v_i,c_j)$ can also be displayed to provide voters with a sense of their relative proximity to different candidates. The exact calculation of such a score is mostly arbitrary. In Smartvote, the Euclidean distance between the voter and candidate is scaled by the maximum possible distance between two answers, as specified in Eq.~\eqref{eq:similarity_score}.
Figure~\ref{fig:top_match_perc} shows that the similarity scores of the best-matching candidate vary by party and are generally quite low, which is in large part a consequence of the curse of dimensionality~\citep{thirey2015distribution}. This disparity could ultimately influence voters from different parties in different ways.

\subsection{Question Designers}
\label{sec:vulnerabilities:questions}

\subsubsection{Question Favoritism (\textcolor{question}{QF})}
Certain questions can significantly benefit specific parties by aligning closely with their popular stances. Figure~\ref{fig:question_favoritism} shows the relative change in party visibility based on the size of alternative questionnaires. These questionnaires consist of a subset of questions from the original set, selected to benefit the respective parties the most during the elections in the state of St. Gallen. With this knowledge, an adversarial question designer could favor questions that benefit their preferred party.

\subsubsection{Question Correlation (\textcolor{question}{QC})}
If a question is advantageous for a particular party, introducing additional questions with answers highly correlated to this question (among voters and candidates) implicitly increases its weight. For instance, asking the negation of a question effectively doubles the original question’s weight. Although this strategy is inherently associated with question favoritism, it has the potential to magnify its impact.

\section{Measuring Robustness}
\label{sec:metrics}

From Table~\ref{tab:dangers_overview}, we note that three high-risk vulnerabilities, namely \textcolor{candidate}{AC}, \textcolor{candidate}{AO}, and \textcolor{platform}{MM}, are highly dependent on the matching method. To assess the impact of matching methods on robustness, we compare the five most commonly used distance functions and two novel proposals using various key robustness metrics. A formal definition of these distance functions is provided in Appendix~\ref{app:metrics:distance}. 

\subsubsection{Party Bias (BIA).} We assess the deviations in party visibility for each matching method relative to the median visibility observed across all other evaluated methods (see Appendix~\ref{app:metrics:BIA} for a detailed discussion). Here we consider the mean absolute deviation (BIA1) and max deviation (BIA2) over the eight largest parties.

\subsubsection{Calibration Potential (CP).}
For each matching method, we repeat the analysis of Figure~\ref{fig:neutral_answer_strategy_recommendation_shares} and measure the average relative visibility gain or loss that results from a party employing the moderate answering strategy (CP-M) or the strong answering strategy (CP-S) weighted by the adjusted voter shares of the parties in the 2023 election (see Table~\ref{tab:parties} in Appendix~\ref{app:smartvote} for the exact values).

\subsubsection{Answer Strength Correlation (ASC).} 
This metric addresses the answer calibration manipulation strategy. It is defined as the Pearson correlation between the answer strength (defined in Eq.~\ref{eq:answer_strength}) and the expectation-normalized visibility of candidates. The expectation-normalized visibility adjusts for the varying number of candidates in each state by multiplying the visibility by the ratio of the number of candidates to the number of available seats in the state, ensuring comparability across different states. To minimize the effectiveness of any answer calibration strategy regarding the answer strength, this metric should ideally be close to zero, indicating no systematic bias toward candidates with moderate or strong answers.

\subsubsection{Gini Coefficient (GIN).}
This metric measures the Gini coefficient of the expectation-normalized visibilities over all candidates, indicating how evenly distributed the recommendations are among them. A Gini coefficient of 0 represents a perfectly even distribution, and a coefficient of 1 indicates a completely uneven distribution. While there is no ideal Gini coefficient for a distance method, and actual election votes are typically less evenly distributed than Smartvote recommendations (see Figure~\ref{fig:gini} in the Appendix), the Gini coefficient offers insight into the differences in recommendation diversity between matching methods.

\subsubsection{Party Match Accuracy (ACC1).}
This metric measures the proportion of voters whose top list recommendation matches their preferred party. As manual accuracy checks are impractical, comparing the voter's stated preferred party with the party recommended by the algorithm is common for assessing the accuracy of VAAs \citep{garziamarschall2014matching}. For Smartvote, which does not directly recommend parties, we use the party from the best-matching list as a proxy. While this metric is appealing for its simplicity, it assumes that voters know the party that best represents them, which may not always be true.

\subsubsection{Normalized Party Rank (ACC2).}
This metric provides deeper insight into the rankings of lists associated with voters’ preferred parties. It measures the average normalized rank of the top list for the preferred party, with normalization adjusting for the number of lists per state. A normalized rank of 0 means the list is recommended first, while a value of 1 means it is recommended last.

\subsubsection{Strong Disagreement Accuracy (ACC3).}
This metric measures the disagreements between voters and their recommended candidates. However, it specifically focuses on questions that voters weighted more strongly, indicating their greater importance. This metric should ideally be low, as voters likely expect their recommended candidates to align with them on these high-priority questions.

\begin{figure}[t]
  \centering
  \includegraphics[width=1\linewidth]{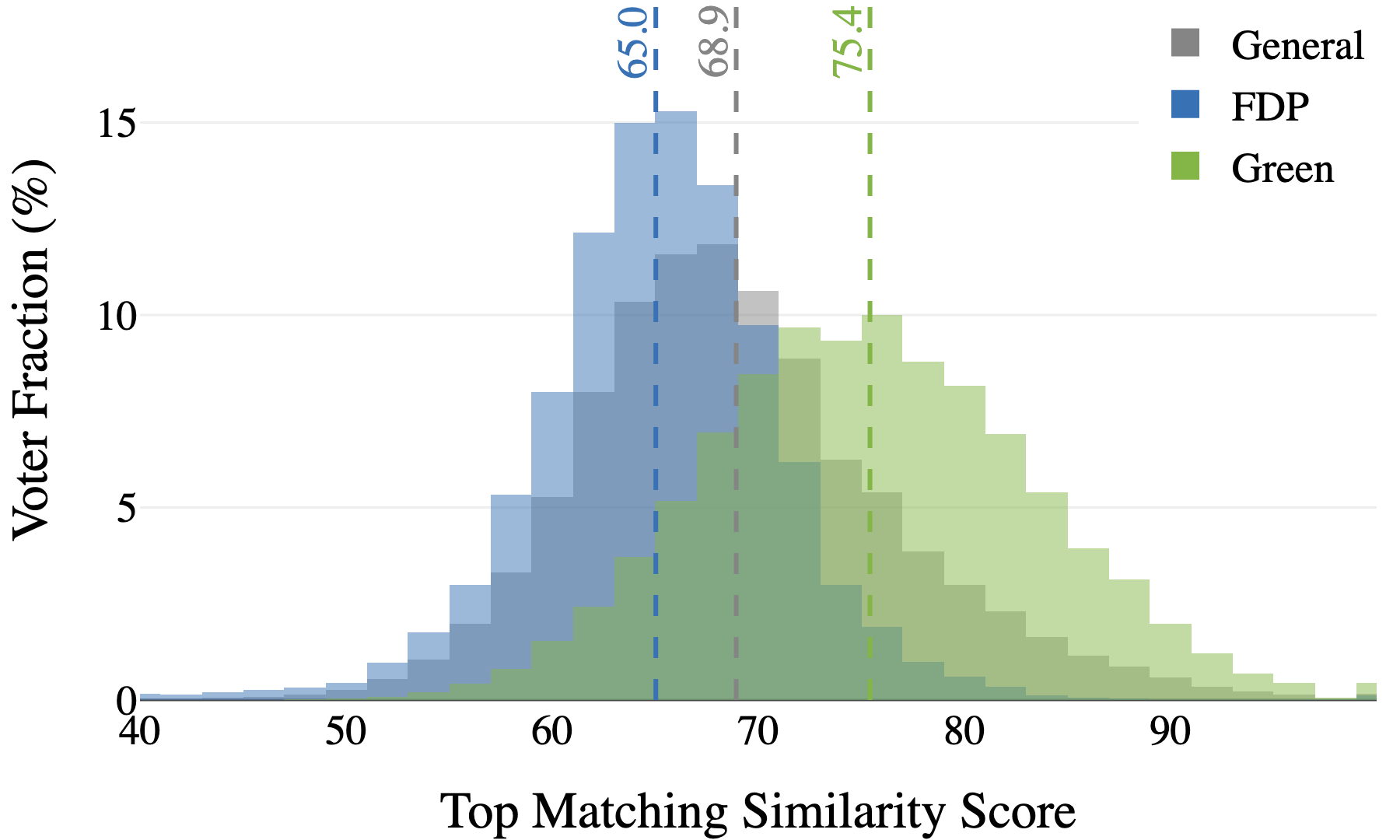}
  \caption{Distribution of similarity scores between voters and their top matching candidate, with colored histograms isolating voters whose top candidate is from a specific party. This histogram reveals that the matching percentages vary significantly based on the party of the top matching candidate. It also shows that for many voters, their top matching candidate is surprisingly low  (below 70\%).
  }
\label{fig:top_match_perc}
\end{figure}

\begin{figure}[t]
    \centering
    \includegraphics[width=1\linewidth]{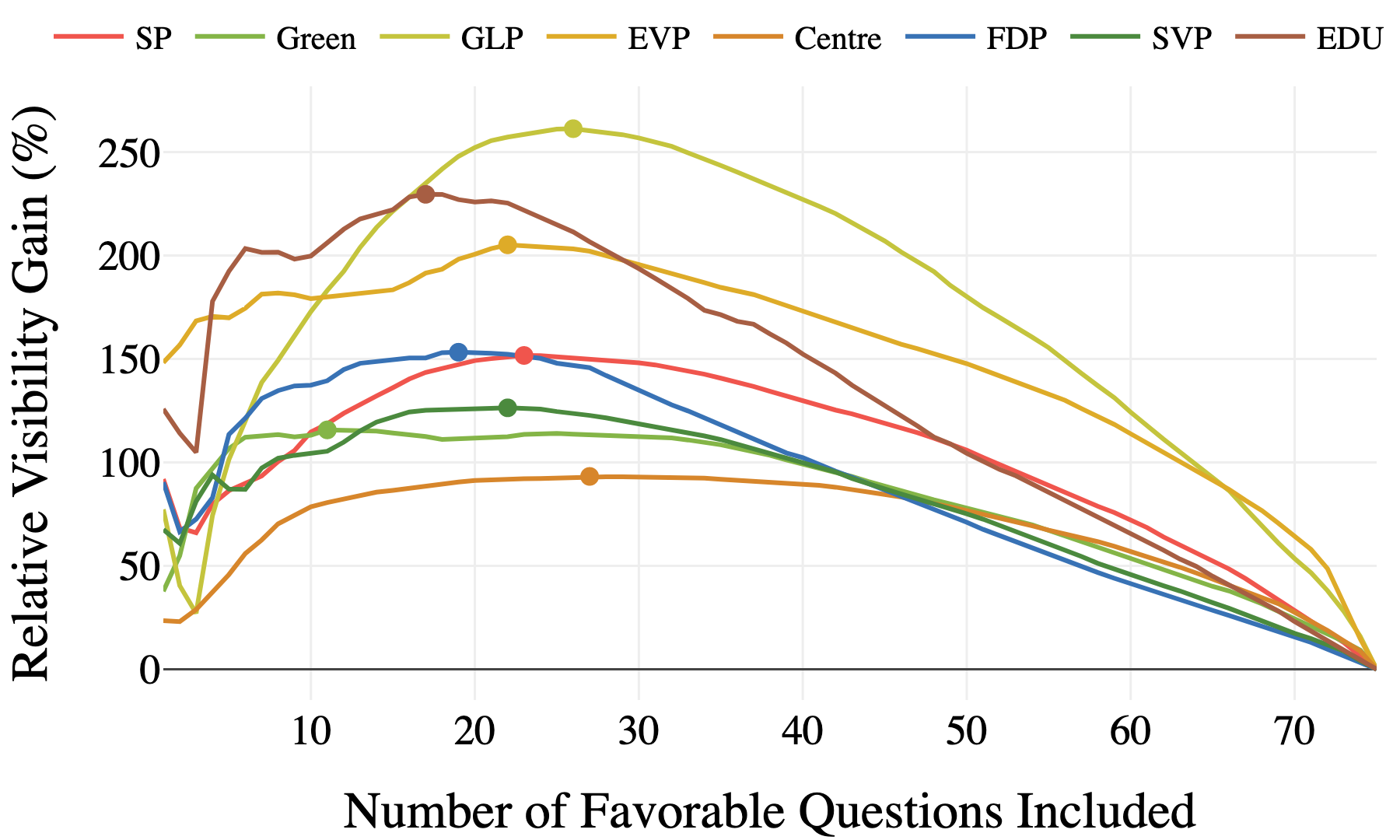}
    \caption{Relative visibility gain for each party using a set of greedily selected optimal questions to generate voting advice in the state of St. Gallen. Each line represents a political party and shows its increase in visibility as more and more favorable questions are included, compared to the baseline scenario with the full questionnaire. Circles indicate each party's maximum attainable visibility (e.g., when only choosing the best-aligned 12 questions, the Green party can increase its visibility by 120\%).}
    \label{fig:question_favoritism}
\end{figure}

\begin{table*}[t]
    \small
    \centering
        \begin{tabularx}{\textwidth}{p{2.3cm}|
        >{\centering\arraybackslash}X
        >{\centering\arraybackslash}p{2.2cm}
        >{\centering\arraybackslash}X
        >{\centering\arraybackslash}X
        >{\centering\arraybackslash}X
        >{\centering\arraybackslash}p{0.6cm}|
        >{\centering\arraybackslash}X
        >{\centering\arraybackslash}X
        >{\centering\arraybackslash}X|
        >{\centering\arraybackslash}p{2.0cm}}
            \toprule
            Distance Function  & BIA1  & BIA2  & CP-M & CP-S  & ASC  & GIN  & ACC1 & ACC2  & ACC3 & Used By \\
            (See App.~\ref{app:metrics:distance}) & $|$\faArrowDown$|$ & $|$\faArrowDown$|$ & \faArrowDown & \faArrowDown & $|$\faArrowDown$|$ &  & \faArrowUp &  \faArrowDown & \faArrowDown & \\
            \midrule
            L2              & 23.0\% & +40.7\% (EVP)   & +207\%  & -71\% & \underline{-0.470} & 0.475 & 41.0\%  & 0.103  & \phantom{0}7.8\% & \textit{Smartvote} \\
            L1              & 14.3\% & +24.4\% (Centre) & +46\%   & -50\% & -0.280 & 0.373 & \textbf{41.8\%}  & \textbf{0.101} & 11.0\%  & Wahl-O-Mat \\
            Angular          & \phantom{0}4.1\% & -12.2\% (GLP)     & \phantom{0}-27\% & -13\% & 0.190     & 0.349 & 40.2\%& 0.109 & \phantom{0}\textbf{7.0\%}    & - \\
            Agreement Count & \phantom{0}\textbf{3.6\%}    & +7.8\% (SVP)     & \phantom{0}-36\% & \phantom{0}-4\%  & 0.256     & 0.317 & 35.7\%& 0.111 & 15.0\%& Stemwijzer \\
            Mahalanobis   & \underline{29.2\%}& -47.2\% (EDU)     & \underline{+305\%}& -69\% & \textbf{0.044}        & 0.523 & \underline{29.0\%}  & \underline{0.142} & \underline{21.6\%}& - \\
            L1 Bonus      & 15.5\%& -27.1\% (GLP)     & -81\% & \underline{+27\%} & \underline{0.583}     & 0.387 & 37.9\%& 0.109 & 11.3\%& Smartvote (old) \\
            Hybrid        & \phantom{0}5.3\% & -15.7\% (GLP)     & -55\% & -12\% & 0.292     & 0.349 & 40.2\%& 0.106 & 10.1\%& EUVox \\
            \bottomrule
        \end{tabularx}
        \caption{Comparison of alternative distance functions based on various metrics defined in Section~\ref{sec:metrics}. The arrows indicate what is desired from the metric (\faArrowUp: Higher is better, \faArrowDown: Lower is better, $|$\faArrowDown$|$: Closer to 0 is better). The best value for each metric is highlighted in bold, and the worst value is underlined. The GIN metric is purely informational, with no suggestion that higher or lower values are better.  \textit{Smartvote (old)} refers to the Smartvote VAA until 2010. A detailed discussion about the counterintuitive CP-M and ASC value for Mahalanobis is provided in Appendix~\ref{app:metrics:ASCvsCP}.}
        \label{tab:method_comparison}
\end{table*}

\section{Future Work on Mitigation Approaches}
\label{sec:mitigations}
Below, we present a series of possible mitigation strategies, specifying the vulnerabilities they aim to address. We also provide mitigations for \textcolor{platform}{TB}, \textcolor{platform}{QO} and \textcolor{candidate}{LC} in Appendix~\ref{app:mitigations}. We emphasize that these strategies have not been extensively tested and may introduce unintended harms. We introduce them here as a foundation for future work, aiming to facilitate systematic research in this direction. Mitigation strategies currently under Politools review are marked with \faSearch.

\subsubsection{\faSearch\, L1 or Angular instead of L2 (\textcolor{candidate}{AC}, \textcolor{candidate}{AO}, \textcolor{platform}{MM}).}
While each distance metric has its trade-offs, we find in Table~\ref{tab:method_comparison} that L1 and Angular consistently offer better robustness than L2 without sacrificing accuracy. Specifically, L1 outperforms L2 in ACC1 and ACC2, while Angular excels in ACC3 with only minor reductions in ACC1 and ACC2. Therefore, we argue that any of these two methods is a viable robust substitution for L2. Alternatively, the Hybrid method appears to offer strong robustness properties with only a slight decrease in accuracy across all three metrics.

\subsubsection{Lower Expressivity (\textcolor{candidate}{AC}, \textcolor{candidate}{AO}).}
Reducing the number of allowable answers can reduce the impact of many vulnerabilities by limiting opportunities for fine-grained manipulation. Since expressivity is important for voters, it could be reduced specifically for candidates. For example, candidates could be restricted to answering “Yes” or “No” for each question, while voters still have “Rather yes” and “Rather no” as options. This would effectively mitigate the answer calibration strategy, which, based on our subjective assessment in Table~\ref{tab:dangers_overview}, poses the greatest risk.

\subsubsection{\faSearch\, Deal-breaker Filtering (\textcolor{platform}{WS}).}
As demonstrated by the vulnerability to weight selection, voters could easily misunderstand the effect of weighting questions. To address this issue, we propose to allow only the weights to ${W_t=\{0,1,\infty\}}$ for each question $q_t$. Assigning a weight $\infty$ to a question effectively treats it as a deal-breaker \citep{isotalo2021improving}, directly excluding all candidates who answered differently from the voter on that question. To avoid leaving voters without candidates due to excessive filtering, the matching algorithm could consider the number of disagreements on deal-breakers as the primary factor in determining the similarity scores. Alternatively, one could also allow voters to exclude all candidates not aligned with their chosen side of the answer spectrum relative to the neutral response.

\subsubsection{\faSearch\, Selective Answering (\textcolor{question}{QF}, \textcolor{question}{QC}).}
Voters should be informed that answering more questions does not necessarily lead to a more accurate recommendation and may even distort the results. The user interface could instead promote a more selective approach to question selection by each voter.

\subsubsection{Distance to Party Mean (\textcolor{candidate}{AC}, \textcolor{candidate}{AO}).}
Voters often lack tools to assess a candidate's honesty and determine if they have answered truthfully or exploited VAA vulnerabilities to boost their visibility. One solution is to display the distance between each candidate’s answers and their party’s mean answers. A large distance might prompt voters to scrutinize the candidate’s responses more closely. However, this metric would only be a proxy for honesty, as some candidates may naturally deviate from their party's position~\citep{schwarz2010pre}.

\subsubsection{Limiting the Number of Candidates (\textcolor{candidate}{DIV}).}

To prevent parties from disproportionately boosting their visibility by increasing candidate numbers, we propose limiting the number of candidates from the same party that can be recommended to any voter. This limit could be based on the similarity score between the voter's position and each party's average position. For instance, if two parties have the same similarity score with a given voter but one has more candidates, the top $k$ recommendations should be evenly distributed between the parties, minimizing the risk of biased recommendations arising from the diversification strategy.

\subsubsection{Fair Answer Normalization (\textcolor{platform}{SS}).}
To avoid presenting varying similarity scores to voters from different parties (as shown in Figure~\ref{fig:top_match_perc}), we propose normalizing similarity scores relative to the top candidate for each voter (who would always be considered a 100\% match). While this would change the score’s meaning and might reduce its overall usefulness, it would also eliminate bias.

\section{Related Work}
\label{sec:related_work}
VAAs emerged around 30 years ago and have quickly gained popularity since then. \citet{garzia2012under-review} provide a comprehensive overview of existing VAAs, summarized in Table~\ref{tab:vaa_database_overview} in the Appendix. The voter data collected by VAAs are a treasure trove, for political, social, and computer scientists alike. \citet{etter2014mining} for example, extract valuable data on the Swiss political landscape. An extended related work discussion on the influence of VAAs on democratic institutions and their development is detailed in Appendix~\ref{app:related_work}.

\subsubsection{VAAs under Scrutiny.}
\citet{walgrave2009voting} show that the question selection has a substantial impact on the voting advice. %
\citet{louwerse2014designeffects} highlight the significant impact matching methods (mainly L1 and L2) have on recommendations, using \textit{StemWijzer} as an example. We corroborate this finding but crucially demonstrate that these matching methods behave differently in the presence of an adversary.
\citet{van2017curse} critically analyze current methods to visualize aggregate results, and propose a technique based on learned dimensions to correct shortcomings.
Finally, \citet{isotalo2021improving} identifies several issues with Finnish VAAs, including lack of transparency, user interactivity, and problems in statement structure. Our work supports the effectiveness of their suggested filtering method.

\subsubsection{Adversarial Robustness of Recommender Systems.}
Other applications have recognized the importance of adversarial robustness~\citep{hurley2011robustness, tang2019adversarial} and the challenges of questionnaire design~\citep{pasek2010optimizing}. \citet{ovaisi2022rgrecsys} provide a toolkit to compare the robustness of learning-based recommender systems. Given the much stricter requirements of recommender systems for democracy (see Section~\ref{sec:background}), while our introduced metrics apply to all methods, we restrict our evaluation to non-learning-based methods for now. 

\section{Outlook}
\label{sec:outlook}

This study highlights critical vulnerabilities in voting advice applications (VAAs), providing empirical evidence that malicious actors could pose a risk to democratic processes. Crucially, many vulnerabilities also uncover the existence of strong biases in VAAs, even in the absence of adversarial entities. 
We are convinced that VAAs are a highly desirable addition to the political landscape and believe that our proposed comparative metrics and mitigations can help guide future VAA development toward more robust designs.
As VAAs continue to evolve in the era of AI, future work should also aspire to extend our results to other types of political recommender systems that fall outside our formalism.

\section*{Ethical Statement}
\label{sec:ethics}

The dataset has been collected and anonymized by Politools in accordance with the new Swiss Federal Act on Data Protection (nFADP), the Telecommunications Act (TCA), and other applicable data protection regulations~\citep{smartvote2024privacy}.
Further, we strictly follow the platform's terms of use for data. As such, we do not publish results that may be attributed to specific individuals. In accordance with the terms of use for research, the dataset is kept private, and we adhere to established best practices for dealing with sensitive data. While the dataset cannot be made accessible directly, it might be made available to researchers by Politools upon request~\citep{smartvote2024website}. Given access to the data, all numerical results and figures can be easily reproduced using the code in the supplementary material.

In the absence of established Ethics guidelines, we follow Menlo's report on Computer Science research principles~\citep{kenneally2012menlo}.
Publicly disclosing all found vulnerabilities presents a risk, as various actors might benefit from exploiting them. We mitigate these risks by publishing our results after Switzerland's national election, leaving enough time to implement potential mitigation for the 2027 elections. To the best of our knowledge, no countries with popular VAAs that could be affected by our research will hold national elections in the months following the publication of this work. Thus, we believe that this is the right time to shed light on these vulnerabilities.
Overall, we believe that despite some inherent risks, this work will have a clear net positive social impact by providing tools to enhance the robustness of VAAs, and consequently, democracies.

\section*{Acknowledgments}
We thank Michael Erne and Daniel Schwarz from Politools for the uncomplicated and fruitful collaboration, as well as their helpful feedback. Our gratitude extends to Roger Germann, Douglas Orsini-Rosenberg, Leon Plath, and Gina Stoffel, whose Bachelor and Master thesis contributed to our understanding of VAAs and the success of this project. Finally, we thank Judith Beestermöller and Robin Fritsch for their valuable input and guidance. 

\bibliographystyle{named}
\bibliography{bibliography}

\clearpage
\appendix

\section{Smartvote and Swiss Politics}
\label{app:smartvote}

\subsection{Swiss National Elections}
\label{app:smartvote:swisselections}

The Swiss parliament is composed of two chambers: the National Council and the Council of States. Together they hold the legislative power in Switzerland, meaning they are responsible for passing laws and approving the federal budget. The National Council is constituted of 200 seats, which are assigned proportionally to the states' populations (approximately). This proportional distribution of seats to the states ensures that each seat in the National Council represents approximately the same number of voters.

\begin{figure}[h]
    \centering
    \includegraphics[width=\linewidth]{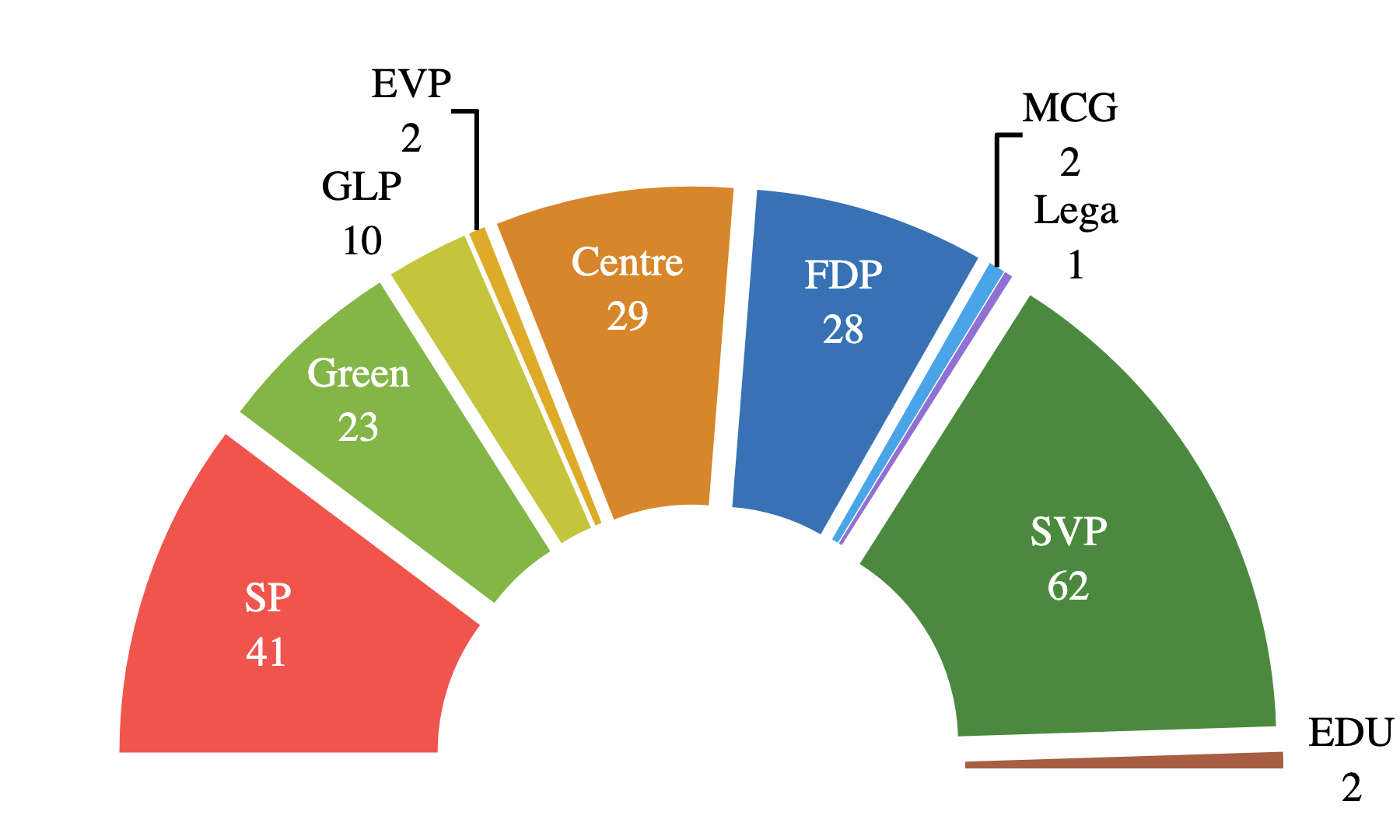}
    \caption{Seat distribution of the National Council after the 2023 elections.}
    \label{fig:seat-distribution-2023}
\end{figure}

The distribution of seats in the National Council following the 2023 elections is illustrated in Figure~\ref{fig:seat-distribution-2023}. The largest party is the Swiss People's Party (SVP), which holds 31\% of the seats, followed by the Social Democratic Party of Switzerland (SP) with 20.5\%. Parliamentary elections are held every four years, and the Swiss parliament is elected on a cantonal basis. This means voters can only cast votes for candidates who stand for election in the same state as their place of residence. The number of candidates a voter can vote for is determined by the number of seats allocated to the state of residence of the voter in the National Council.

In most states, candidates are required to form a coalition, known as a candidate list, to be eligible to stand for election. Voters are then presented with the option of either voting for a predefined list of candidates or composing their custom list of individual candidates. The number of candidates on a list corresponds to the number of seats available in the state. Furthermore, voters have the option to give more weight to specific candidates by placing a candidate's name twice on the list, effectively giving them two votes.
The political landscape of Switzerland's federal government comprises 11 major parties whose name, abbreviation, current number of seats, and vote share in the 2023 National Council elections are summarized in Table~\ref{tab:parties}.

\setcounter{table}{3}
\begin{table}[h!]
    \small
    \centering
    \begin{tabularx}{\linewidth}{p{3.5cm} p{0.5cm} >{\centering\arraybackslash}X >{\centering\arraybackslash}X}
        \toprule
        Party & Code  & \centering Vote Share &  \centering Vote Share (adjusted)\arraybackslash \\
        \midrule
        Swiss People's Party        & SVP   & 27.9\%            & 27.1\% \\
        Social Democratic Party     & SP    & 18.2\%            & 19.1\% \\
        The Centre                  & Centre & 14.3\%            & 14.7\% \\
        FDP. The Liberals           & FDP   & 14.1\%            & 13.8\% \\
        Green Party                 & Green & \phantom{0}9.8\%  & 10.4\% \\
        Green Liberal Party         & GLP   & \phantom{0}7.6\%  & \phantom{0}8.0\% \\
        Evangelical People's Party  & EVP   & \phantom{0}2.0\%  & \phantom{0}2.0\% \\
        Federal Democratic Union    & EDU   & \phantom{0}1.2\%  & \phantom{0}1.1\% \\
        Party of Labour             & PdA   & \phantom{0}0.7\%  & \phantom{0}0.4\% \\
        Lega dei Ticinesi           & Lega  & \phantom{0}0.6\%  & \phantom{0}0.2\% \\
        Geneva Citizens Movement    & MCG   & \phantom{0}0.5\%  & \phantom{0}0.3\% \\
        \bottomrule
    \end{tabularx}
    \caption{Overview of the major parties in the 2023 Swiss National Council elections. The table displays the party names, their codes, their vote shares as a percentage of the total national vote, and their adjusted vote shares (calculated by excluding votes cast for candidates not registered in Smartvote). Tiny parties and candidates without party affiliations are excluded, explaining why percentages may not add up to 100\%.}
    \label{tab:parties}
\end{table}
\setcounter{table}{2}
\begin{table*}[t]
    \centering
    \small
    \renewcommand{\arraystretch}{1.1} %
        \small
        \begin{tabularx}{\textwidth}{p{3.3cm} 
        >{\centering\arraybackslash}p{3cm} 
        >{\centering\arraybackslash}p{1.5cm}
        >{\centering\arraybackslash}p{1cm}
        >{\centering\arraybackslash}X 
        >{\centering\arraybackslash}X 
        p{2cm}}
        \toprule
        VAA Name                    & Region                                    & Type                                          & $|A_t|$   & $W_t$                             & Distance Metric \\ \midrule
        Smartvote                   & \worldflag[length=0.4cm, width=0.4cm]{CH} 
                                      \worldflag[length=0.6cm, width=0.4cm]{LU}  
                                      \worldflag[length=0.6cm, width=0.4cm]{AT}
                                      \worldflag[length=0.6cm, width=0.4cm]{AU}
                                                                                & \icon{\faUser} \icon{\faList} \icon{}         & 4, 5, 7   & $\{0, 0.5, 1, 2\}$                         & L2 \\ 
        Wahl-O-Mat                  & \worldflag[length=0.6cm, width=0.4cm]{DE} & \icon{} \icon{} \icon{\faFlag}                & 3         & $\{0, 1, 2\}$                              & L1 \\ 
        StemWijzer (VoteMatch)      & \worldflag[length=0.6cm, width=0.4cm]{NL}  
                                      \worldflag[length=0.6cm, width=0.4cm]{BG} 
                                      \worldflag[length=0.6cm, width=0.4cm]{CZ} & \icon{} \icon{} \icon{\faFlag}                & 3         & $\{0, 1, 2\}$                            & Agreement Count \\ 
        Wahlkabine                  & \worldflag[length=0.6cm, width=0.4cm]{AT} & \icon{} \icon{} \icon{\faFlag}                & 2         & $\{0, 1, 2, 3, 4, 5, 6, 7, 8, 9\}$           & Agreement Count \\ 
        VoteCompass                 & \worldflag[length=0.6cm, width=0.4cm]{US}   
                                      \worldflag[length=0.6cm, width=0.4cm]{AU} & \icon{\faUser} \icon{} \icon{\faFlag}         & 5         & $\{0, 1\}$                                 & L1 \\
        Volkskabin, Politikkabine   & \worldflag[length=0.6cm, width=0.4cm]{HU} 
                                      \worldflag[length=0.6cm, width=0.4cm]{AT} & \icon{\faUser} \icon{} \icon{\faFlag}         & 2         & $\{0, 0.5, 1, 2\}$\footnote{This VAA slightly deviate from our framework as candidates also have the option to assign weights questions.}  & Agreement Count (adj.) \\
        Manobalsas                  & \worldflag[length=0.6cm, width=0.4cm]{LT} & \icon{\faUser} \icon{} \icon{\faFlag}         & 5         & $\{0, 1\}$                                 & L2 (low-dim.) \\
        EU\&I (Euandi)              & \worldflag[length=0.6cm, width=0.4cm]{EU} & \icon{} \icon{} \icon{\faFlag}                & 5         & $\{0, 0.5, 1, 2\}$                           & L1 \\
        EUVox 2014                  & \worldflag[length=0.6cm, width=0.4cm]{EU} & \icon{} \icon{} \icon{\faFlag}                & 3         & $\{0, 1\}$                               & Hybrid \\
        iSideWith                   & \worldflag[length=0.6cm, width=0.4cm]{US} 
                                      \worldflag[length=0.6cm, width=0.4cm]{CA} 
                                      \worldflag[length=0.6cm, width=0.4cm]{MX}  
                                      \worldflag[length=0.6cm, width=0.4cm]{NZ} 
                                      \worldflag[length=0.6cm, width=0.4cm]{KR}  
                                      \worldflag[length=0.6cm, width=0.4cm]{IE} 
                                      \worldflag[length=0.6cm, width=0.4cm]{PL} & \icon{\faUser} \icon{} \icon{}                & 2         & 5 weights (undisclosed)                        & Agreement Count (adj.) \\
        \bottomrule
        \end{tabularx}    
        \caption{The table includes the name, region (flags), type of recommendation (\icon{\faUser}: candidates, \icon{\faList}: lists, \icon{\faFlag}: parties), number of answer options, question weights, and distance metric used by some of the most popular VAAs globally.}
        \label{tab:vaa_database_overview}
\end{table*}
\setcounter{table}{4}

\subsection{Questionnaire Design}
\label{app:smartvote:questionnaire}

The questionnaire is the core element for capturing political positions and is newly created for each election. The creation process starts months before the elections and involves the collection of question proposals from official party or government websites, media analysis, and public submissions. Over 1,500 proposals were gathered for the 2019 Swiss federal elections. The Smartvote team independently refines these proposals into a 45 to 75-question survey, ensuring clarity, neutrality, and relevance to current and future political discussions. Finally, the draft questionnaire is quality-checked by experts from the scientific community and selected Smartvote users.

The final questionnaire consists of 75 questions split into three question types: 60 policy questions with ${A_t = \{0, 25, 75, 100\}}$, 7 value questions with ${A_t = \{0, 17, 33, 50, 67, 83, 100\}}$, and 8 budget questions with ${A_t = \{0, 25, 50, 75, 100\}}$. Policy questions capture positions on specific political issues (e.g., \emph{Do you support an increase in the retirement age?}). Value questions capture agreement or disagreement with broad political principles (e.g., \emph{What is your position on the following statement: “In the long term, everyone benefits from a free market economy.”?}). Budget questions capture how the spending on key federal budget items should be adjusted (e.g., \emph{Should the federal government spend more or less in the area of social services?}).

Voters also have the option to complete a rapid version of the questionnaire, which consists of the 30 policy questions deemed most important by the questionnaire designers, as an alternative to the full 75-question deluxe version.

\subsection{Smartvote Usage}
\label{app:smartvote:usage}

Before each National Council election cycle, candidates are invited to participate by filling out a detailed questionnaire. This invitation process typically starts a few months before the election, allowing candidates enough time to complete their profiles. All candidates are required to answer all questions in the questionnaire. Candidates then have time to complete their profiles until the questionnaire is made public to the voters. After the publications candidates are generally not allowed to adjust their answers anymore. In individual, justified exceptional cases, however, corrections have been made after publication in the case of clear errors.

Once the candidate profiles are finalized, Smartvote is made available to the public. Voters can then access the platform, fill out the questionnaire, and receive voting advice based on the candidate's responses. Additionally, voters have the option to provide metadata such as demographic details, political preferences, and their interest in politics before completing the questionnaire. Voters are free to use Smartvote as often as they like until the election date, allowing them to make informed decisions based on the most up-to-date information available. It is important to note that the voters using Smartvote represent a biased sample of the Swiss voting population. Our analysis indicates that Smartvote users are generally younger, more likely to reside in urban areas and have a preference for left-leaning parties compared to the overall voting population. This bias may influence the results and should be considered when interpreting the data.

\begin{figure*}[t]
\begin{subfigure}{0.47\linewidth}
    \centering
    \includegraphics[width=\linewidth]{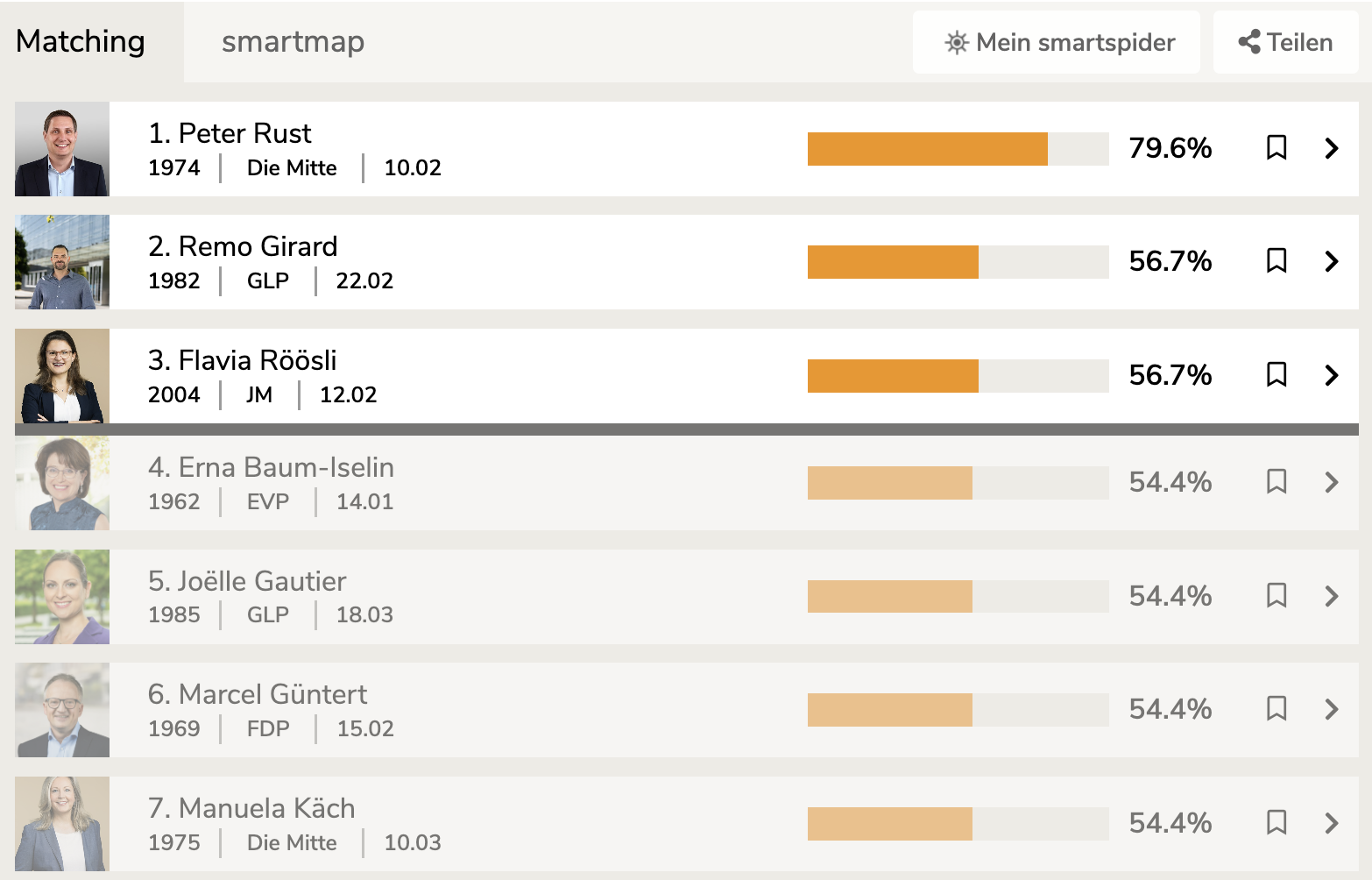}
    \caption{Smartvote UI for candidate recommendations.}
    \label{fig:candidate_recommendations_example}
\end{subfigure}
\hfill
\begin{subfigure}{0.47\linewidth}
    \centering
    \includegraphics[width=\linewidth]{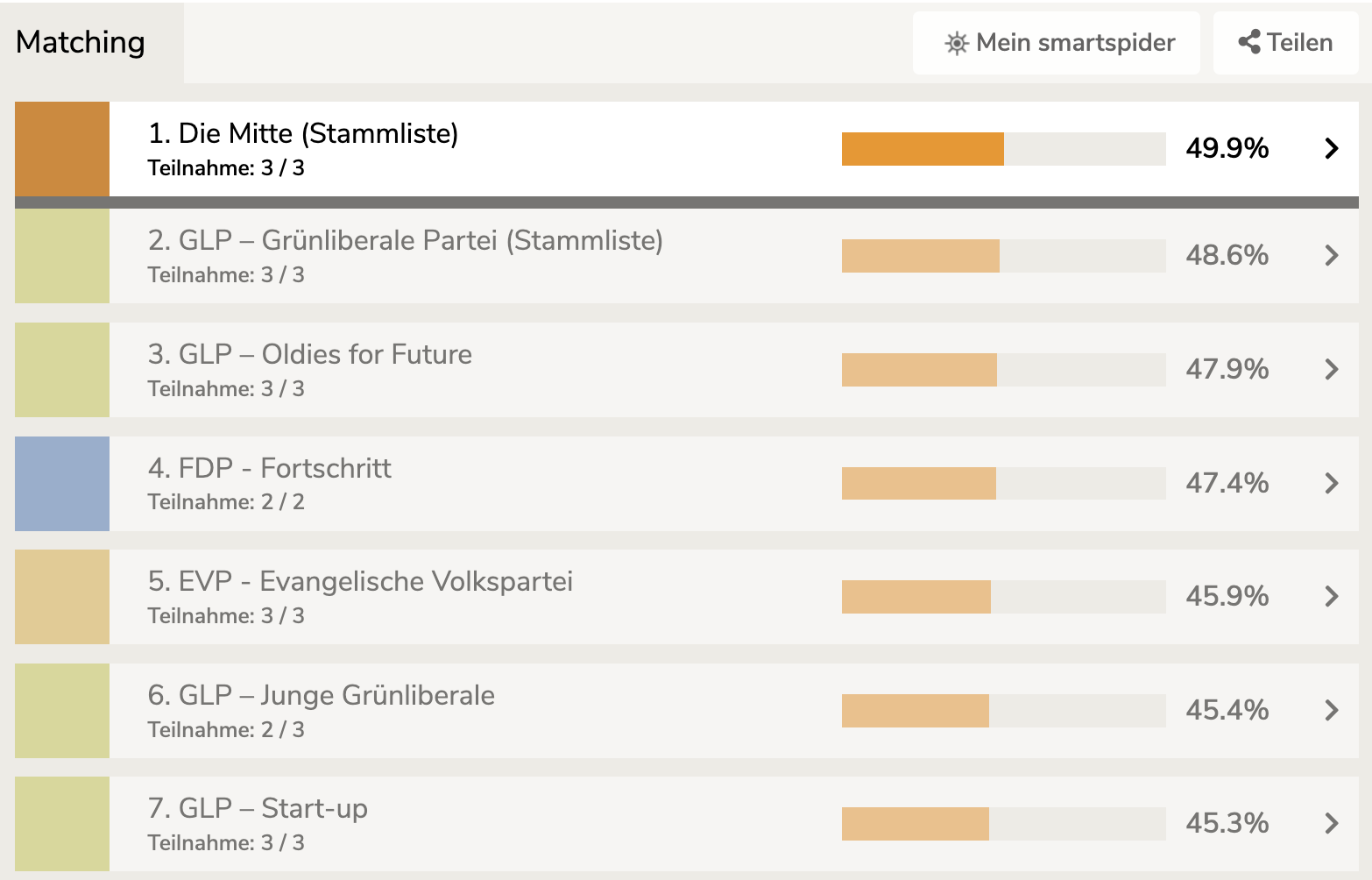}
    \caption{Smartvote UI for list recommendations.}
    \label{fig:list_recommendations_example}
\end{subfigure}
\caption{Screenshots of candidate and list recommendations on Smartvote for a voter in the state of Zug. Candidates and lists are ranked by their similarity score in descending order. Given that the state of Zug has three seats available in the National Council (which is also the list size for that state), the top three candidates and the first list are prominently displayed. This example also highlights how Smartvote resolves ties between candidates with identical similarity scores using their last names.}
\end{figure*}
\subsection{Smartvote UI}
\label{app:smartvote:smartvote_ui}

Figure~\ref{fig:candidate_recommendations_example} illustrates how candidate recommendations are displayed on Smartvote. It shows how the top-ranked candidates are prominently featured based on the number of seats available in the state. This also demonstrates our definition of candidate visibility, determined by the proportion of voters for whom a candidate is prominently displayed. Similarly, Figure~\ref{fig:list_recommendations_example} illustrates how list recommendations are presented to voters, effectively demonstrating the concept of list visibility, which is defined as the proportion of voters for whom a given list is prominently displayed.

\subsection{Limitations}
\label{app:smartvote:limitations}

To validate our implementation of the matching algorithm used by Smartvote and to ensure the integrity of the dataset, we tried to reproduce the recommendations from the 2023 Smartvote elections dataset, representing the exact recommendations voters were given on the platform. Despite our best efforts, we encountered certain limitations. A small number of candidates deleted their profiles after the publication of Smartvote, which meant that recommendations involving them were not fully reproducible. Additionally, 17 candidates were allowed to correct their responses after the publication due to clear errors, making the recommendations involving them before the corrections not fully reproducible.

\section{Dataset (continued)}
\label{app:dataset}

\subsection{Preprocessing \& Cleaning}
\label{app:dataset:cleaning}

To enhance the quality of our data and prepare it for analysis, we conducted thorough preprocessing on both the voter and candidate datasets from both elections.
For the voter datasets, we first addressed corrupt entries by identifying and removing those lacking essential information, such as the election ID.
To ensure the reliability of our analysis, we excluded recommendations where voters answered fewer than 15 questions, which accounted for less than 1.3\% of the dataset.
Further refinement involved filtering out recommendations both before the questionnaire went live and after the election date. This step aimed to exclude recommendations intended for testing purposes and those no longer relevant to voting decisions.
To streamline our dataset, we implemented a deduplication process. We retained only the most recent recommendations with the highest number of answered questions for each unique voter ID, ensuring data consistency.
Lastly, to eliminate recommendations resulting from random clicking behavior, we filtered out those with more than 14 consecutive identical answers, which accounted for about 0.1\% of the dataset.
As for the candidate datasets, our preprocessing primarily involved filtering out candidates who did not participate in Smartvote as the dataset included all candidates up for election.
As part of the preprocessing, we ensured that all variables used in our analysis contained valid and sensible values for all voters and candidates. Additionally, we created a consolidated party variable by merging the youth wings of political parties with their respective main parties, consistent with the preferred party variable provided by users. For instance, JUSO (Young Socialists) was merged with SP (Social Democratic Party), and JSVP (Young Swiss People's Party) was combined with SVP (Swiss People's Party). Table \ref{tab:party_combinations} provides an overview of how youth parties were mapped to the main parties.

\begin{table}[h]
   \small
    \centering
    \begin{tabularx}{\linewidth}{X X}
        \toprule
        Youth Party & Main Party \\
        \midrule
        JUSO                   & SP                      \\
        JG                     & Green                   \\
        JGLP                   & GLP                     \\
        JEVP                   & EVP                     \\
        JM                     & Centre                  \\
        JFS                    & FDP                     \\
        JSVP                   & SVP                     \\
        \bottomrule
    \end{tabularx}
    \caption{Party combinations showing how youth parties are combined into main parties.}
    \label{tab:party_combinations}
\end{table}

\subsection{Exploratory Data Analysis}
\label{app:dataset:exploratory}

\begin{table}[h]
   \small
    \centering
    \begin{tabularx}{\linewidth}{X p{1.5cm} p{1.5cm} X X X}
        \toprule
        Dataset & \# Rec. & $N_v$ & $N_c$ & Lists & $N_q$ \\
        \midrule
        2019 & 427,572 & 389,881 & 3,926 & 508 & 75 \\
        2023 & 1,662,683 & 485,838 & 4,983 & 623 & 75 \\
        \bottomrule
    \end{tabularx}
    \caption{Dataset overview for the 2019 and 2023 National Council elections. The table includes the number of voting recommendations before cleaning (\# Rec.), the number of recommendations after cleaning (taken as the number of unique voters $N_v$), the number of participating candidates ($N_c$), the number of lists, and the number of questions in each dataset ($N_q$).}
    \label{tab:dataset_overview}
\end{table}

To gain a deeper understanding of the datasets and provide intuition to the reader, we performed an exploratory data analysis built in large part around a Principal Component Analysis (PCA) on the voter and candidate answer profiles. These analyses help to uncover underlying patterns and structures in the data, which are crucial for evaluating and improving the matching algorithm.

\begin{figure*}
\begin{subfigure}{0.47\linewidth}
    \centering
    \includegraphics[width=\linewidth]{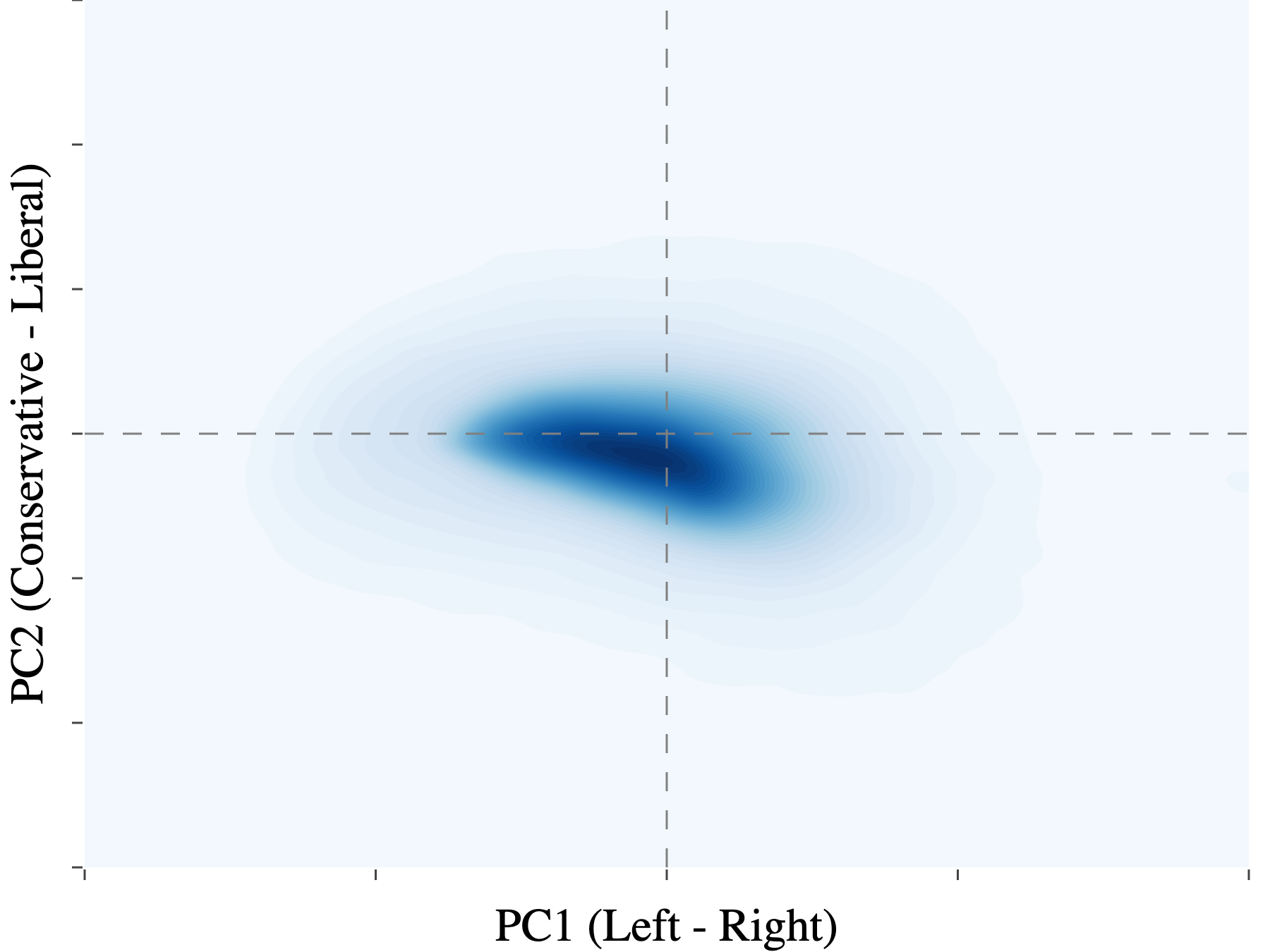}
    \caption{Voters' answer distribution.}
    \label{fig:voter_density}
\end{subfigure}
\hfill
\begin{subfigure}{0.47\linewidth}
    \centering
    \includegraphics[width=\linewidth]{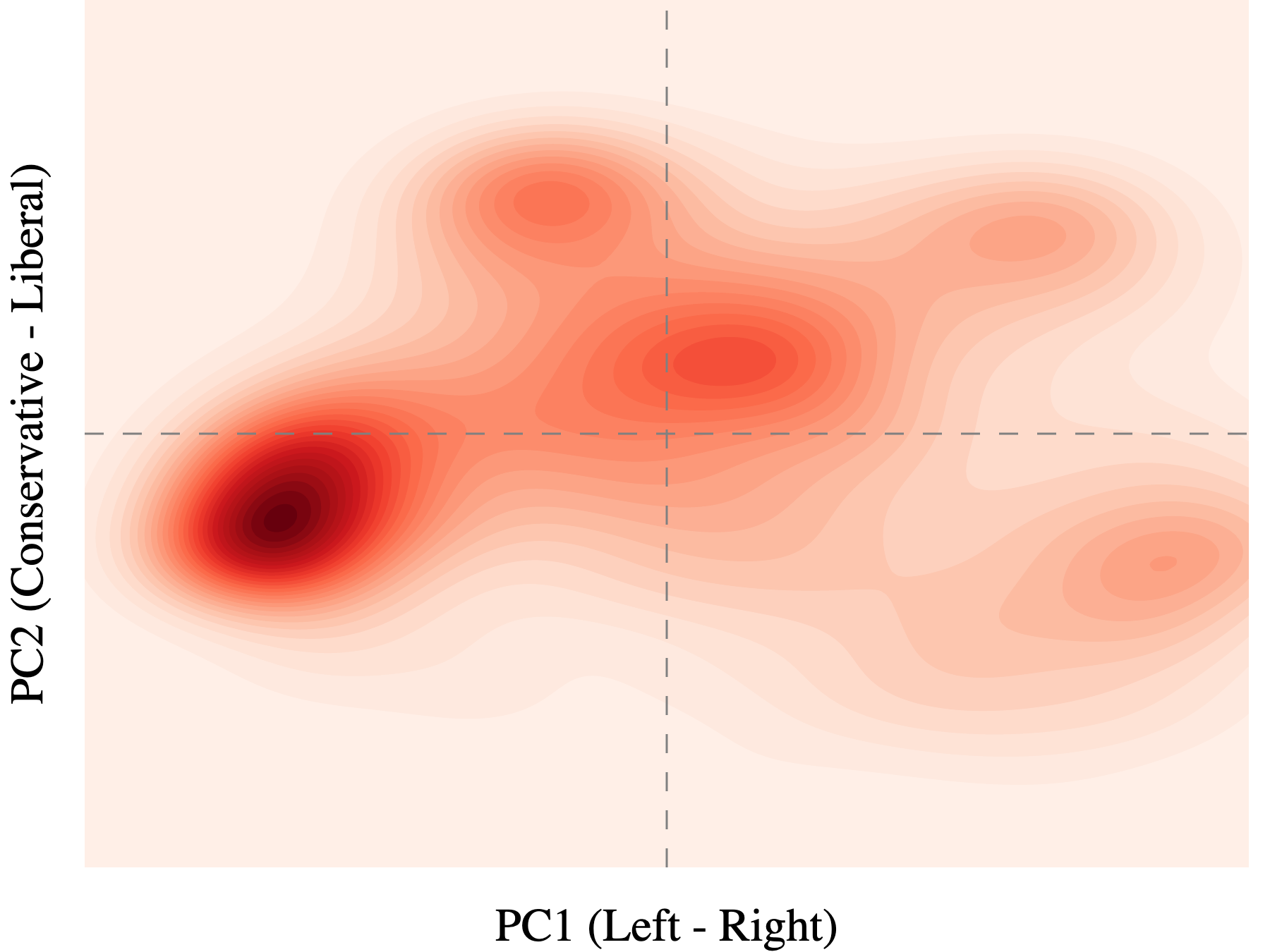}
    \caption{Candidates' answer distribution.}
    \label{fig:candidate_density}
\end{subfigure}
\caption{Density plots of voters and candidates in the 2D political space for the 2023 election, showing the distribution of their answers along the two principal components dimensions of the candidate answers. These dimensions can be interpreted as economic Left-Right and social Conservative-Liberal \citep{garziamarschall2014matching, smartvote2024website}.}
\end{figure*}

One of the key analyses involves visualizing the density distributions of voters and candidates in a two-dimensional political space defined by the two first principal components of the candidate answer profiles. These dimensions can be roughly interpreted as Left-Right and Conservative-Liberal. The first plot (Figure~\ref{fig:voter_density}) depicts the voter density for the 2023 election, showing how voters are distributed across this political landscape. The second plot (Figure~\ref{fig:candidate_density}) illustrates the candidate density, highlighting where candidates position themselves in the same political space. Comparing these plots reveals the alignment or gaps between voter preferences and candidate positions, providing valuable insights into the effectiveness of the VAA's recommendations.

\begin{figure}[h]
    \centering
    \includegraphics[width=\linewidth]{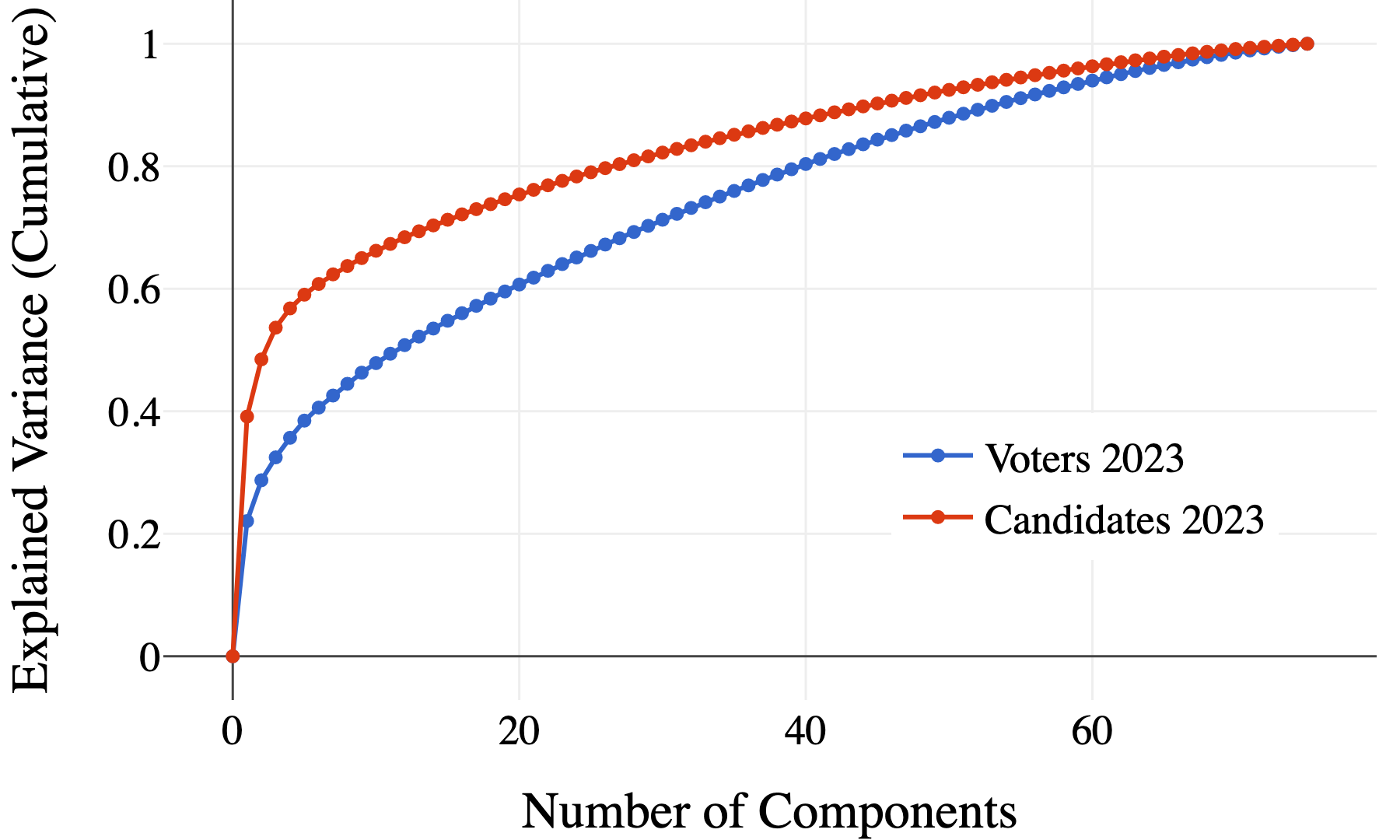}
    \caption{PCA explained variance for voter and candidate answer profiles in the 2023 dataset. The plot shows the cumulative explained variance as a function of the number of principal components for both voters and candidates.}
    \label{fig:explained_variance}
\end{figure}

\begin{figure}[h]
    \centering
    \includegraphics[width=\linewidth]{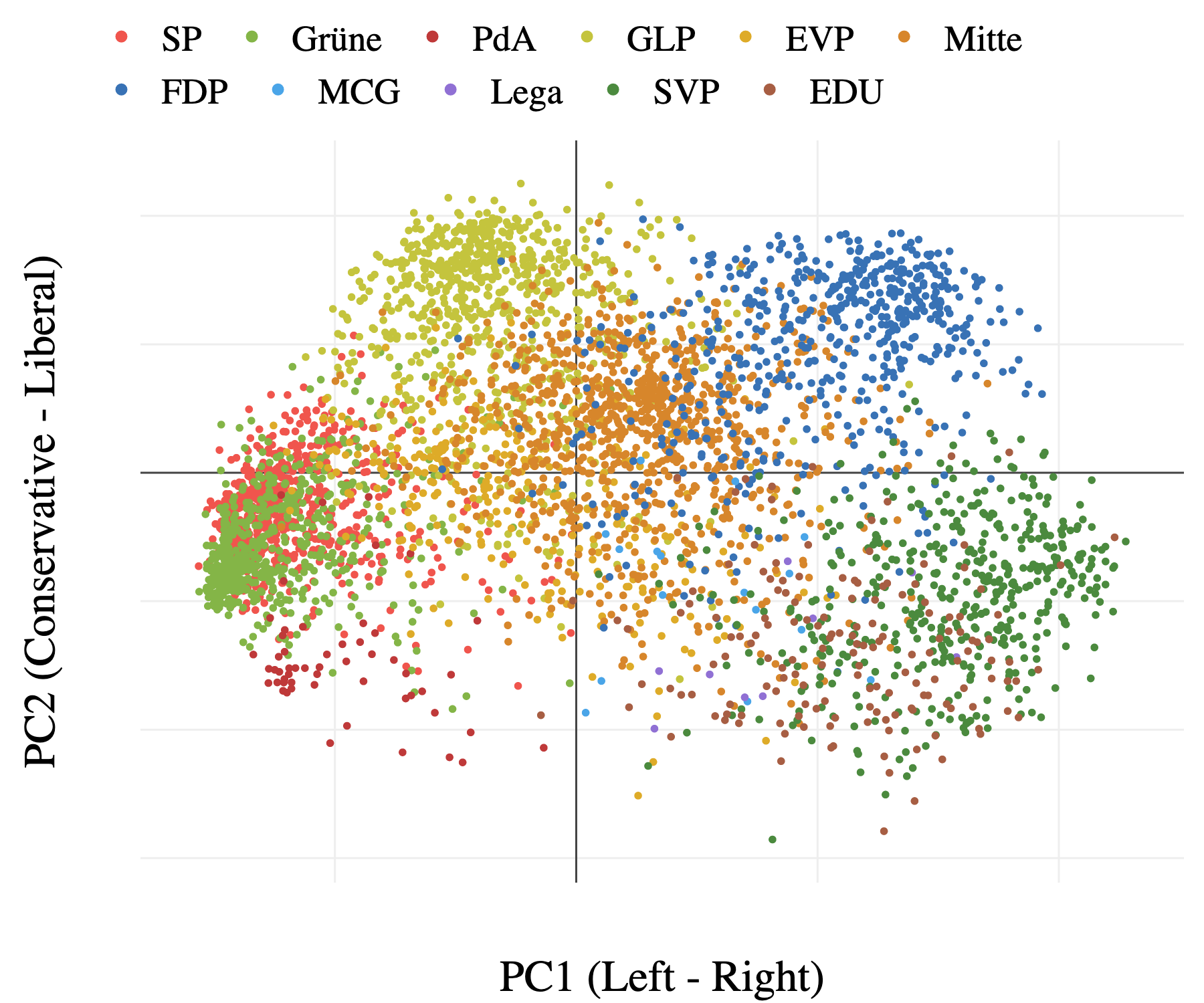}
    \caption{PCA plot of candidate positions by party in the 2023 dataset, highlighting candidate clustering and each party's political orientation.}
    \label{fig:candidate_positions}
\end{figure}

Additionally, Figure~\ref{fig:explained_variance} shows the cumulative explained variance of answers as a function of the number of principal components, for both voters and candidates. From this plot, we observe that while the first three principal components capture more than 50\% of the variance in the candidate data, they only capture around 30\% of the variance in the voter data. This suggests that candidate responses are more structured and consistent, likely due to alignment with party platforms, whereas voters' responses are more varied, reflecting a broader range of opinions and requiring more principal components to capture the same variance. This observation aligns with findings by \citet{kleinnijenhuis2016dimensionality}, who noted that voters' policy preferences are not strongly structured and cannot be easily captured by low-dimensional spatial models. 
We also visualize the positions of individual candidates in a two-dimensional space. Figure~\ref{fig:candidate_positions} displays the PCA plot of the candidates' positions, colored by their respective parties. This visualization highlights the distribution and clustering of candidates along the primary axes of political orientation, which can be interpreted as Left-Right and Conservative-Liberal \citep{garziamarschall2014matching, smartvote2024website}. From the PCA plot, we can observe distinct clustering patterns among different parties. For instance, the SP and PdA parties, represented in red, are predominantly situated in the left-wing quadrant. Conversely, the SVP, shown in green, occupies the right-wing, conservative area of the plot. 
The FDP party, colored in blue, is more dispersed but generally aligns with liberal positions. This differentiation underscores the varying political ideologies and strategies of each party.

\section{Vulnerabilities (continued)}
\label{app:vulnerabilities}
\subsubsection{Supplementary Material} 
\begin{figure}[h]
    \centering
    \includegraphics[width=0.8\linewidth]{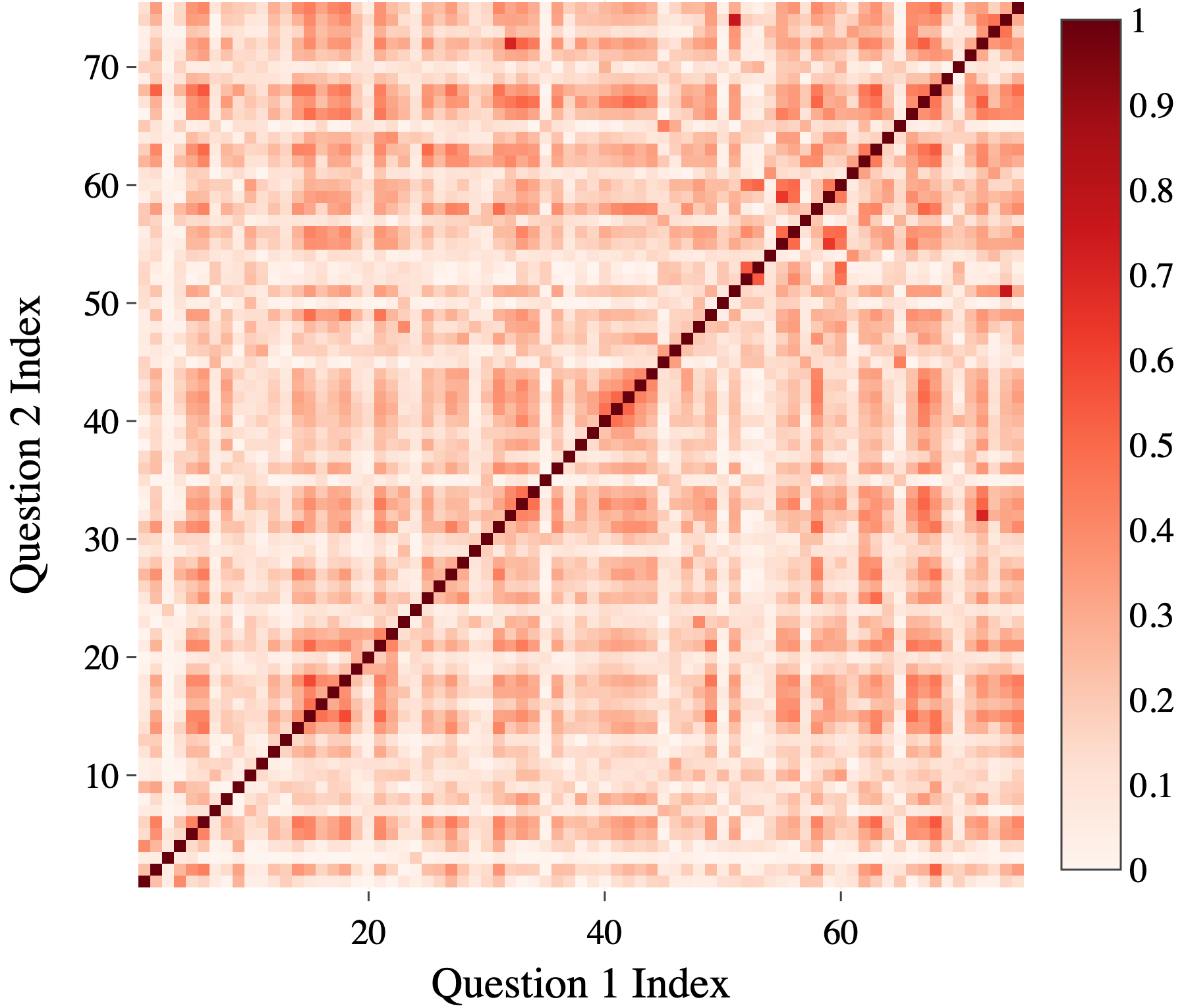}
    \caption{Absolute correlation between question responses in the voter dataset. The heatmap reveals clusters of highly correlated questions ($\approx 0.8$), suggesting that topics covered by these questions are implicitly weighted more heavily for the recommendation calculation. This provides an advantage to parties that are typically favored by these questions.}
    \label{fig:correlation_heatmap}
\end{figure}

As additional material for the \textcolor{candidate}{AO} vulnerability, Table~\ref{tab:optimization_overview} provides the visibility of the crafted candidates for all states and different optimization strategies. Concerning the \textcolor{platform}{MM} vulnerability, Table~\ref{tab:method_party_visibilities} provided detailed visibility measurement for each party using several matching methods. Lastly, Figure~\ref{fig:correlation_heatmap} displays the correlation between the voters' answers, supporting the fact that \textcolor{question}{QC} can have a non-negligible impact on recommendations.
We now expose three additional vulnerabilities, namely \textcolor{candidate}{LC}, \textcolor{platform}{QO} and \textcolor{platform}{TB}.

\begin{table*}[t]
\centering
\renewcommand{\arraystretch}{1.2} %
\small
\begin{tabularx}{\linewidth}{p{3.5cm}
>{\raggedleft\arraybackslash}p{1cm}
>{\raggedleft\arraybackslash}p{1cm}
>{\raggedleft\arraybackslash}p{1cm}
>{\centering\arraybackslash}X
>{\centering\arraybackslash}X
>{\centering\arraybackslash}X}
\toprule
State (canton) & \centering $N_v$ & \centering $N_c$ & \centering Seats & \centering Highest Visibility & \centering Visibility  Optimized 1\% Data & \centering Visibility Optimized 100\% Data \arraybackslash\\
\midrule
Zurich                  & 104,826 & 1,029 & 36  & 24.77\% & 51.70\% & \textbf{53.43\%} \\
Bern                    & 89,378  & 685   & 24  & 29.87\% & 50.66\% & \textbf{53.71\%} \\
Aargau                  & 47,442  & 568   & 16  & 20.44\% & 51.29\% & 53.99\% \\
Lucerne                 & 32,378  & 329   & 9   & 31.12\% & 49.70\% & 53.66\% \\
St. Gallen              & 27,421  & 288   & 12  & 32.19\% & 52.55\% & \textbf{61.71\%} \\
Vaud                    & 25,591  & 337   & 19  & 37.66\% & 54.51\% & 64.92\% \\
Valais                  & 20,586  & 199   & 8   & 25.70\% & 55.27\% & 61.82\% \\
Fribourg                & 18,982  & 137   & 7   & 35.07\% & 59.83\% & 63.84\% \\
Solothurn               & 16,364  & 163   & 6   & 19.29\% & 52.71\% & 59.84\% \\
Basel-Landschaft        & 15,872  & 163   & 7   & 20.60\% & 51.08\% & 58.05\% \\
Thurgau                 & 14,644  & 187   & 6   & 16.58\% & 44.52\% & 54.73\% \\
Basel-Stadt             & 13,233  & 107   & 4   & 18.07\% & 46.91\% & 57.10\% \\
Graubünden              & 10,666  & 109   & 5   & 36.27\% & 55.93\% & 61.92\% \\
Geneva                  & 9,892   & 217   & 12  & 35.05\% & 43.13\% & 63.92\% \\
Schwyz                  & 8,593   & 98    & 4   & 25.54\% & 44.13\% & 58.01\% \\
Zug                     & 7,287   & 84    & 3   & 22.73\% & 34.49\% & 55.11\% \\
Neuchâtel               & 6,689   & 56    & 4   & 46.17\% & 32.74\% & 70.82\% \\
Ticino                  & 5,023   & 144   & 8   & 29.94\% & 25.22\% & 58.21\% \\
Schaffhausen            & 3,257   & 36    & 2   & 33.93\% & 41.63\% & 64.05\% \\
Jura                    & 3,039   & 34    & 2   & 16.75\% & 33.30\% & 60.09\% \\
Appenzell Ausserrhoden  & 1,263   & 2     & 1   & 79.41\% & 53.13\% & 96.75\% \\
Glarus                  & 892     & 3     & 1   & 51.79\% & 21.19\% & 76.01\% \\
Nidwalden               & 806     & 3     & 1   & 72.58\% & 51.36\% & 88.46\% \\
Uri                     & 699     & 2     & 1   & 72.53\% & 66.95\% & 95.42\% \\
Obwalden                & 662     & 2     & 1   & 75.83\% & 77.49\% & 96.22\% \\
Appenzell Innerrhoden   & 353     & 1     & 1   & 100.0\% & 79.89\% & 100.0\% \\
\bottomrule
\end{tabularx}
\caption{List of states along with their corresponding number of unique voters ($N_v$), registered candidates ($N_c$), and the number of available seats for the 2023 National Council elections. The \emph{highest visibility} denotes the most visible candidate in the dataset. The last two columns show the visibility of candidates generated using the \emph{simulated annealing} method \citep{kirkpatrick1983optimization} with the first $1\%$ of the data and with $100\%$ of the data, respectively. The values in \textbf{bold} correspond to the red dots shown in Figure~\ref{fig:optimization} of the main paper.}
\label{tab:optimization_overview}
\end{table*}

\begin{table*}[h!]
    \centering
    \renewcommand{\arraystretch}{1.3} %
    \small
    \begin{tabularx}{\linewidth}{p{3cm}
    >{\centering\arraybackslash}X
    >{\centering\arraybackslash}X
    >{\centering\arraybackslash}X
    >{\centering\arraybackslash}X
    >{\centering\arraybackslash}X
    >{\centering\arraybackslash}X
    >{\centering\arraybackslash}X
    >{\centering\arraybackslash}X
    >{\centering\arraybackslash}X
    >{\centering\arraybackslash}X
    >{\centering\arraybackslash}X}
        \toprule
        Distance Metric & SP & Green & PdA & GLP & EVP & Centre & FDP & MCG & Lega & SVP & EDU \\
        \midrule
        Vote Share & 18.2\% & 9.8\% & 0.7\% & 7.6\% & 2.0\% & 14.3\% & 14.1\% & 0.5\% & 0.6\% & 27.9\% & 1.2\% \\
        Preferred Party & 27.60\% & 14.28\% & 0.74\% & 18.27\% & 1.92\% & 11.01\% & 13.52\% & 0.02\% & 0.07\% & 9.79\% & 0.57\% \\
        \midrule
        L2 & \negColor{100}14.79\% & \negColor{22}12.29\% & \negColor{12}0.69\% & \negColor{38}15.07\% & \posColor{100}8.49\% & \posColor{100}21.88\% & \negColor{68}7.97\% & \posColor{64}0.14\% & \negColor{41}0.06\% & \negColor{62}6.43\% & \posColor{100}2.85\% \\
        L1 & \negColor{85}16.73\% & \negColor{1}14.21\% & \posColor{20}0.83\% & \negColor{49}14.2\% & \posColor{85}7.47\% & \posColor{80}19.69\% & \negColor{69}7.86\% & \posColor{75}0.16\% & \negColor{24}0.06\% & \negColor{47}7.27\% & \posColor{97}2.78\% \\
        Agreement Count & \negColor{68}18.84\% & \posColor{25}16.54\% & \posColor{46}0.95\% & \negColor{63}12.97\% & \posColor{63}6.07\% & \posColor{47}16.1\% & \negColor{70}7.79\% & \posColor{88}0.19\% & \negColor{40}0.06\% & \negColor{27}8.32\% & \posColor{94}2.72\% \\
        Angular & \negColor{64}19.38\% & \posColor{34}17.29\% & \posColor{100}1.2\% & \negColor{76}11.92\% & \posColor{69}6.46\% & \posColor{44}15.83\% & \negColor{65}8.2\% & \posColor{88}0.19\% & \negColor{13}0.07\% & \negColor{28}8.26\% & \posColor{93}2.7\% \\
        Mahalanobis & \negColor{35}23.17\% & \posColor{100}23.26\% & \negColor{35}0.58\% & \negColor{30}15.75\% & \posColor{53}5.4\% & \posColor{34}14.7\% & \negColor{100}5.37\% & \posColor{19}0.05\% & \negColor{100}0.04\% & \negColor{100}4.38\% & \posColor{38}1.43\% \\
        L1 Bonus & \negColor{45}21.78\% & \posColor{65}20.09\% & \posColor{96}1.17\% & \negColor{100}9.9\% & \posColor{45}4.87\% & \posColor{19}13.06\% & \negColor{67}8.08\% & \posColor{100}0.21\% & \negColor{22}0.07\% & \negColor{15}8.97\% & \posColor{87}2.56\% \\
        Hybrid & \negColor{61}19.74\% & \posColor{44}18.25\% & \posColor{82}1.12\% & \negColor{82}11.44\% & \posColor{62}6.0\% & \posColor{44}15.82\% & \negColor{67}8.07\% & \posColor{88}0.19\% & \negColor{13}0.07\% & \negColor{30}8.16\% & \posColor{90}2.62\% \\
        \bottomrule
    \end{tabularx}
    \caption{Party visibilities when using different matching methods. The first two rows serve as references: the first row displays the parties’ vote shares in the 2023 Swiss National Council elections, while the second row shows the distribution of preferred parties as indicated by voters. The cell colors indicate deviations from these preferences: higher values are green and lower values are red, with the saturation representing the magnitude of the deviation.}
    \label{tab:method_party_visibilities}
\end{table*}

\subsubsection{List Centralization (\textcolor{candidate}{LC})}
\begin{figure}[t]
    \centering
    \includegraphics[width=\linewidth]{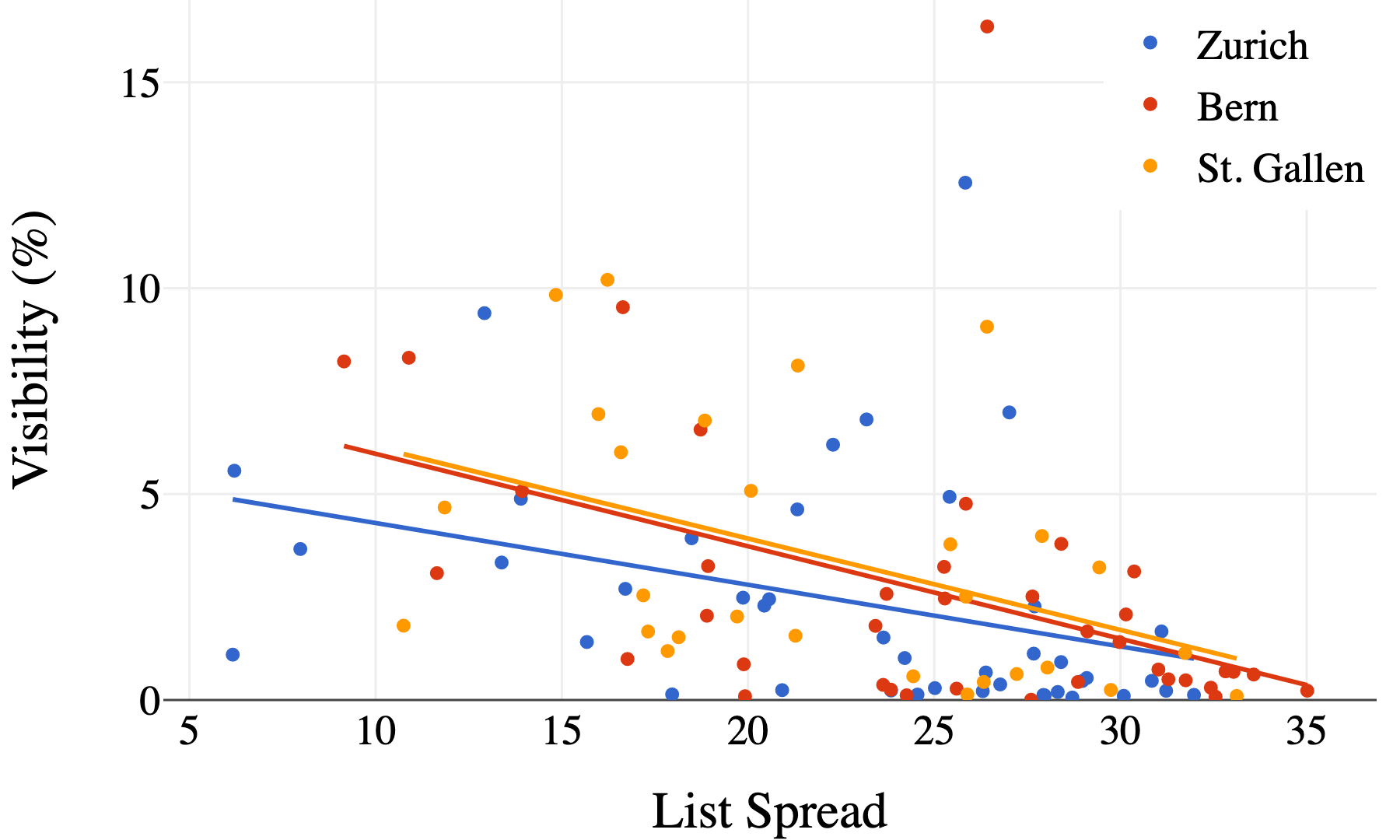}
    \caption{Relationship between the spread and visibility of lists in the states of Zurich, Bern, and St. Gallen. The list spread is defined as the average (over questions) of the answers' standard deviation (across the candidates on the list).}
    \label{fig:list_centralization_correlation}
\end{figure}

This danger pertains to the safety of the list recommendation functionality of VAAs. We show that there exists a bias toward favoring lists that group many candidates with similar answer vectors compared to lists with more diversity. Figure~\ref{fig:list_centralization_correlation} quantitatively demonstrates the impact of this bias for some states. If parties are limited in their number of lists, this creates a trade-off with the diversification strategy (see Section~\ref{sec:vulnerabilities:candidates} in the main paper). However, in Smartvote, parties are allowed to create many lists, making both strategies exploitable simultaneously.

\subsubsection{Tie-breaking (\textcolor{platform}{TB})}
\begin{figure}[t]
    \centering
    \includegraphics[width=1\linewidth]{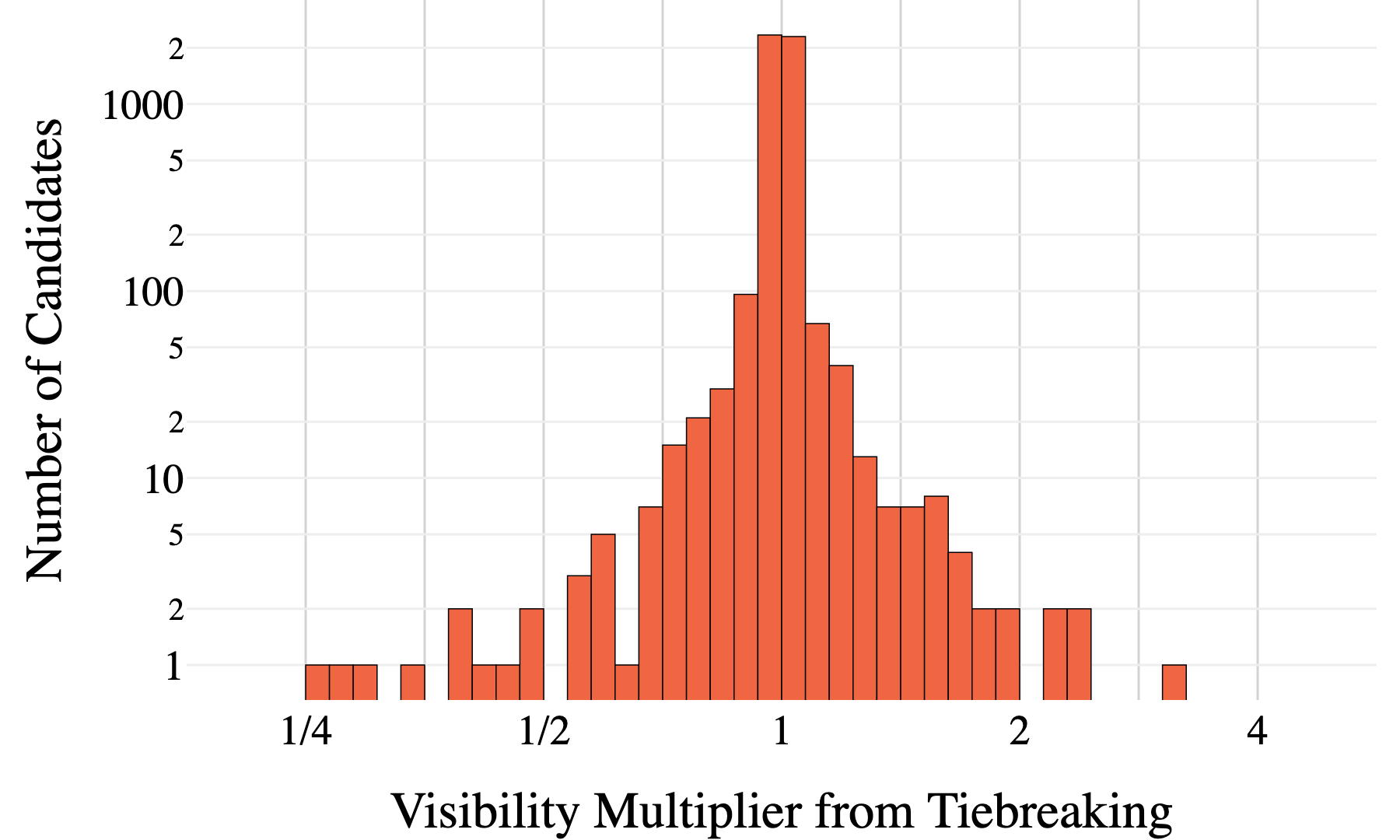}
    \caption{Distribution of relative changes in the visibility of candidates when switching from a fair proportional distribution of ties to Smartvote's tie-breaking method. The x-axis represents the relative visibility change, while the logarithmic y-axis indicates the fraction of candidates affected.}
    \label{fig:tiebreaking_candidates}
\end{figure}

With the extensive number of voters and candidates using Smartvote, and considering that some voters answer only a limited number of questions, multiple candidates may end up at the same distance from a given voter. In such cases, Smartvote resolves ties by using the alphabetical order of the candidates' last names. This approach is problematic, as it systematically favors candidates with last names starting with letters from the beginning of the alphabet over those with last names starting with letters from the end. Figure~\ref{fig:tiebreaking_candidates} shows how the visibility of candidates differs between a fair distribution of ties and the Smartvote tie-breaking, revealing that certain candidates are significantly affected by this tie-breaking methodology.

\subsubsection{Question Ordering (\textcolor{platform}{QO})}
\begin{figure}[t]
  \centering
    \centering
    \includegraphics[width=1\linewidth]{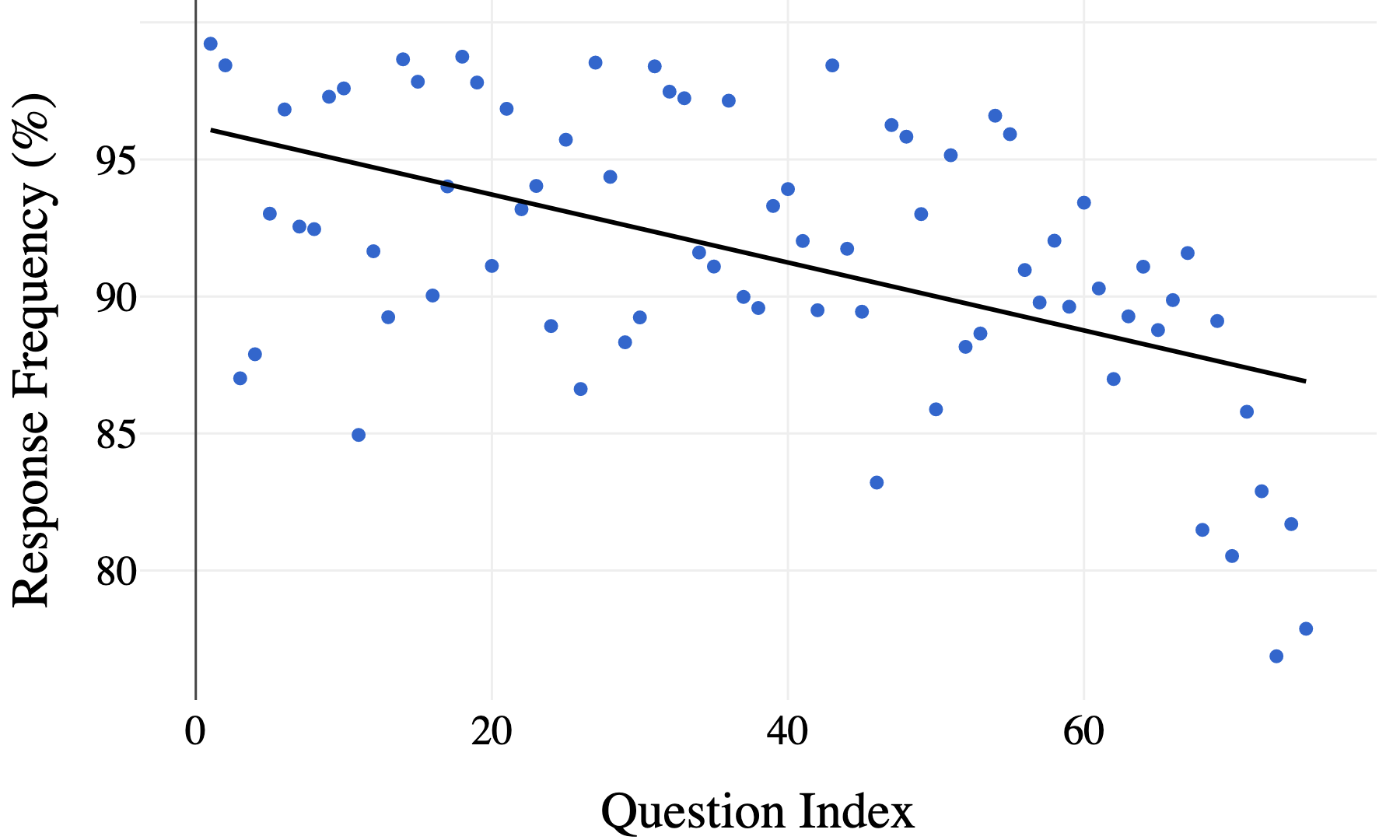}
    \caption{Relationship between response frequency and question position in the deluxe questionnaire (all 75 questions). The black line represents an Ordinary Least Squares (OLS) trend line with equation ${f(t)=96\% - t\cdot 0.12\%}$, where $t$ is the index of the question. The plot shows a pattern where later questions get answered less frequently.}
    \label{fig:response_frequency}
\end{figure}

\begin{figure}[t]
    \centering
    \includegraphics[width=1\linewidth]{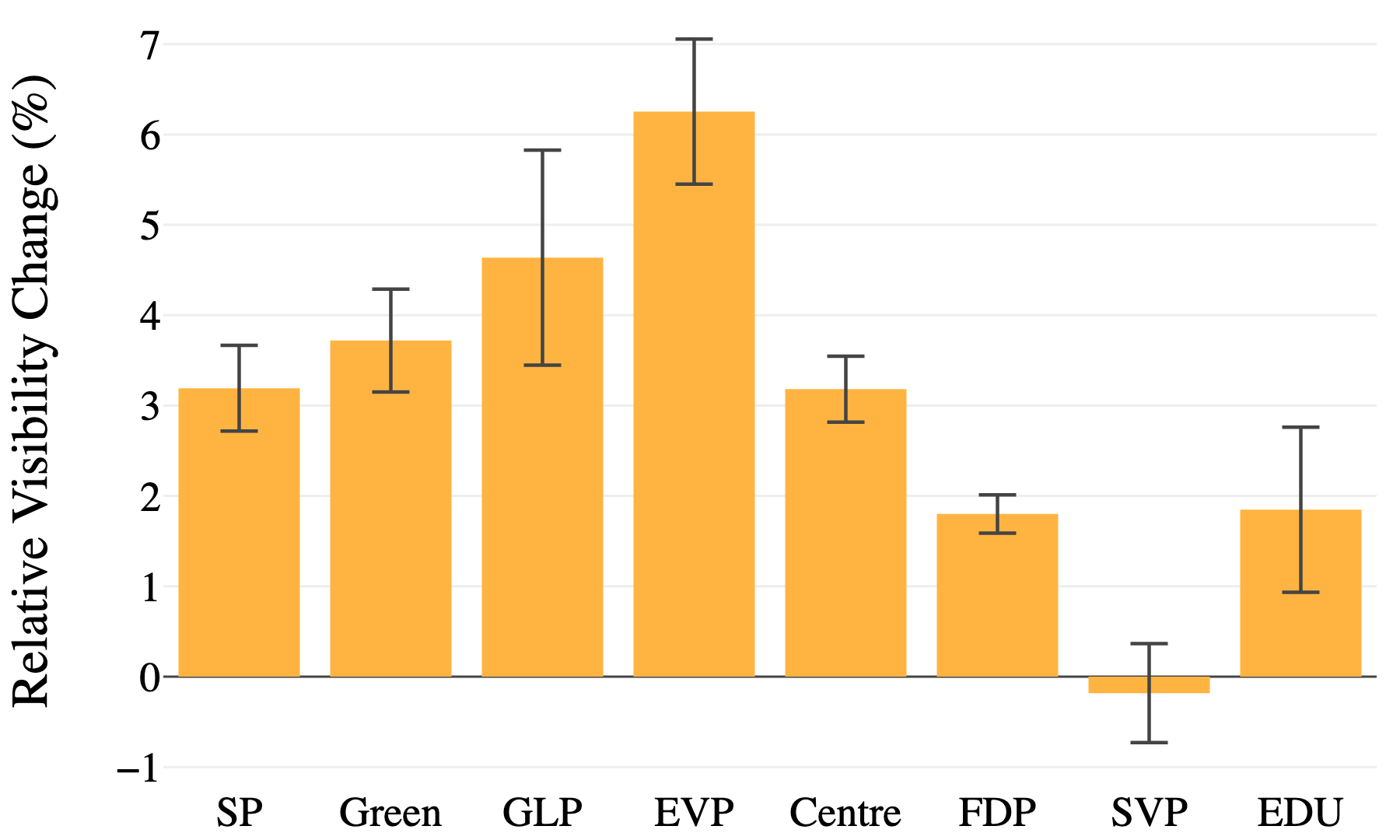}
    \caption{Relative visibility change for each political party resulting from ordering the questionnaire according to the most favorable questions for each party. The bars represent the average visibility change across 10 randomized trials, with error bars indicating the standard deviation. The figure highlights how certain parties, like EVP, could potentially benefit from favorable question orderings.}
    \label{fig:question_ordering}
\end{figure}

The order in which questions are presented can introduce biases such as the primacy effect \citep{asch1946forming}, the priming effect \citep{kuzyakov2000review}, and survey fatigue \citep{jeong2023exhaustive}. The primacy effect indicates that early questions tend to be given more weight, while the priming effect suggests that responses may vary depending on the order of questions. Finally, survey fatigue indicates that later questions are often skipped or answered less thoughtfully, as illustrated in Figure~\ref{fig:response_frequency}. These effects could easily be exploited by an adversarial platform designer. 
To quantitatively understand the potential impact of such a strategy, we perform the following analysis. First, for each $q_t\in Q$, we compute the fraction $f_t$ of answer vectors for which $q_t$ was answered (the dots on Figure~\ref{fig:response_frequency}) and we fit a function $f(t)$ to these values (the line on Figure~\ref{fig:response_frequency}). Then, for each party $p$, we rearranged all complete answer vectors (i.e., the answer vectors of voters in $V^c=\{v_i\mid\mathbf{w}_{i,t}\neq 0, \forall t\}$) following the order shown in Figure~\ref{fig:question_favoritism}. Let $\mathbf{v}^p_i$ be the answer vector of voter $v_i$ rearranged according to $p$'s favorable ordering. Then, for each $v_i\in V^c$, we drop with probability $1-f(t)$ the answer to question $t$ in $\mathbf{v}^p_i$ (i.e., by setting $\mathbf{w}_{i,t}=0$). Finally, we compare the visibility gain of each party with respect to the party's popularity resulting from the original ordering, only considering voters in $V^c$. The results of this analysis are displayed in Figure~\ref{fig:question_ordering}.

\section{Measuring Robustness (continued)}
\subsection{Party Bias (BIA)}
\label{app:metrics:BIA}

To quantify the bias of a distance function, we evaluate relative changes in party visibilities when using this distance function compared to the median party visibility across all other evaluated distance functions. BIA1 represents the average of the absolute values of these relative changes in visibility for the 8 largest parties, ensuring that smaller parties' noisy relative changes do not disproportionally affect the metric. BIA2 captures the maximum absolute value of relative change across the 8 largest parties, highlighting the most significant deviation.

We define the visibility of a party \( p \) when using a specific distance function \( d \) as \( \nu(p \mid P, d) \). Let \( \mathcal{D} \) represent the set of all evaluated distance functions, which includes L2, L1, Agreement Count, Angular, Mahalanobis, L1 Bonus, and Hybrid. Let \( P_8 \) represent the set of the 8 largest parties, which includes SP, Green, GLP, EVP, Centre, FDP, SVP, and EDU. BIA1 and BIA2 are then defined as follows:

\begin{align*}
\text{Median}(p \mid d) & = \text{med}\{\nu(p \mid P, d') \mid d' \in \mathcal{D} \setminus \{d\}\},
\\
\Delta_{\text{rel}}(p, d) & = \frac{\nu(p \mid P, d) - \text{Median}(p \mid d)}{\text{Median}(p \mid d)},
\\
p_{\text{max}} & = \arg\max_{p \in P_8} \left|\Delta_{\text{rel}}(p, d)\right|,
\\
\text{BIA1}(d) & = \frac{1}{\left|P_8\right|} \sum_{p \in P_8} \left|\Delta_{\text{rel}}(p, d)\right|,
\\
\text{BIA2}(d) & = \Delta_{\text{rel}}(p_{\text{max}}, d).
\end{align*}
Table \ref{tab:method_party_visibilities} provides a detailed overview of how party visibilities change based on the matching methods used for calculating recommendations. When comparing these party visibilities with the references, it’s important to note that the Smartvote user base is inherently biased, with only about 85\% of candidates participating in Smartvote and only around 20\% of eligible voters using the platform, not all of whom disclosed their preferred party. A key indicator of this bias is the notable difference between election vote shares and the distribution of preferred parties among Smartvote users.

\subsection{Answer Calibration Metrics (ASC, CP)}
\label{app:metrics:ASCvsCP}
To assess the vulnerability of each matching method to an answer calibration strategy, we employed the Answer Strength Correlation (ASC) and Calibration Potential (CP) metrics. These metrics generally align, meaning that if one suggests a matching method is vulnerable, the other typically indicates the same. However, a notable exception is the Mahalanobis matching method. As shown in Table \ref{tab:method_comparison}, the ASC for Mahalanobis is very low at 0.044, indicating a weak correlation between candidates’ answer strengths and their expectation-normalized visibilities. This suggests that an answer calibration strategy would likely be ineffective. In contrast, the CP metric reveals that parties could increase their visibility by over 300\% when all candidates used the moderate answering strategy. Although this result may initially appear counterintuitive, further analysis offers an explanation: When analyzing the ASC in scenarios where a party’s candidates adopt a moderate answering strategy, the correlation becomes quite negative, with values dropping below -0.6, varying by party. We hypothesize that this is due to the alteration of the precision matrix (inverse covariance matrix), which plays a crucial role in calculating the Mahalanobis distance. This alteration leads to candidates using the neutral answering strategy receiving a significantly increased number of recommendations.

\subsection{Distance Metrics}
\label{app:metrics:distance}
\begin{figure*}[t!]
    \centering
    \begin{subfigure}{0.24\linewidth}  %
        \centering
        \includegraphics[width=\linewidth]{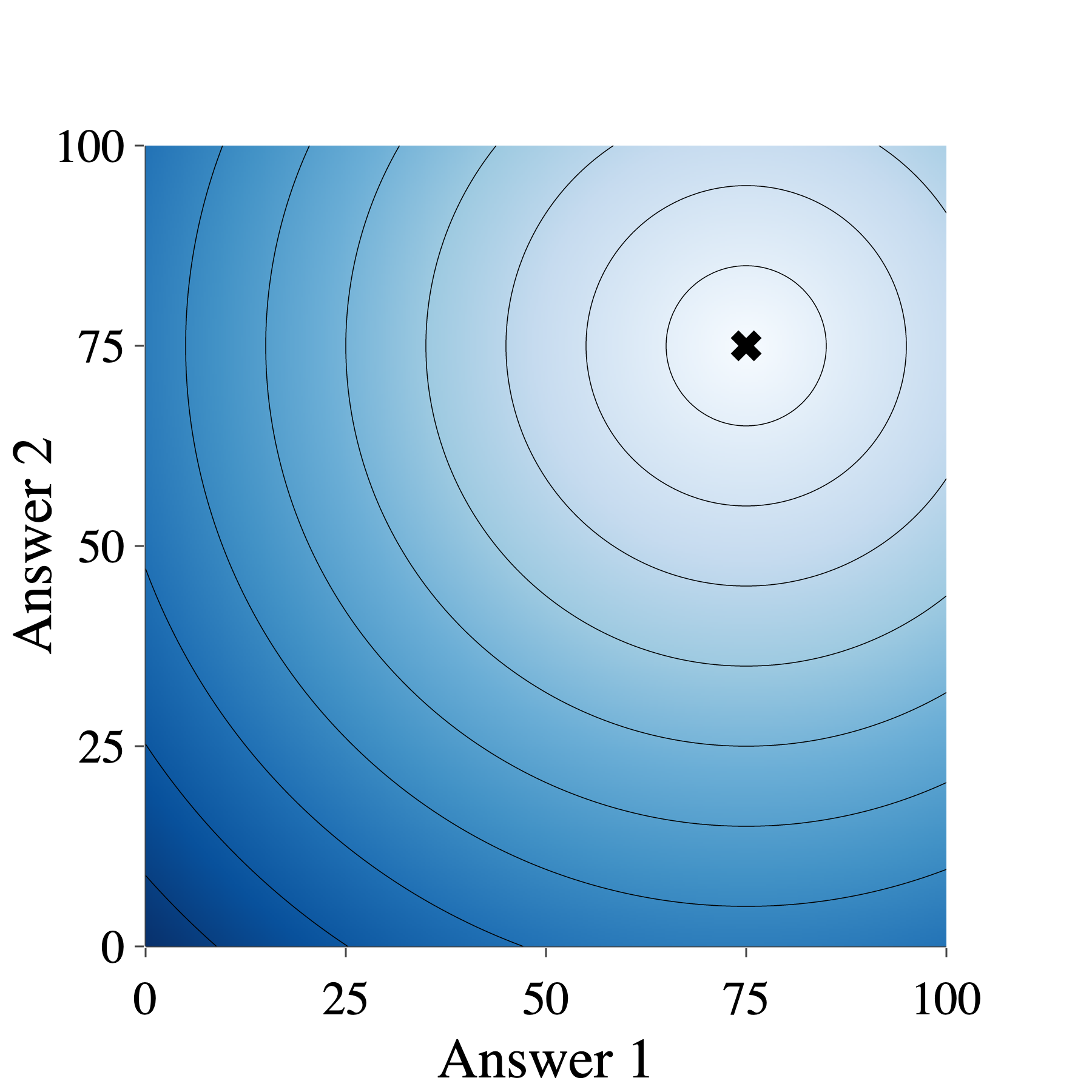}
        \caption{L2}
        \label{fig:L2}
    \end{subfigure}
    \hfill
    \begin{subfigure}{0.24\linewidth}  %
        \centering
        \includegraphics[width=\linewidth]{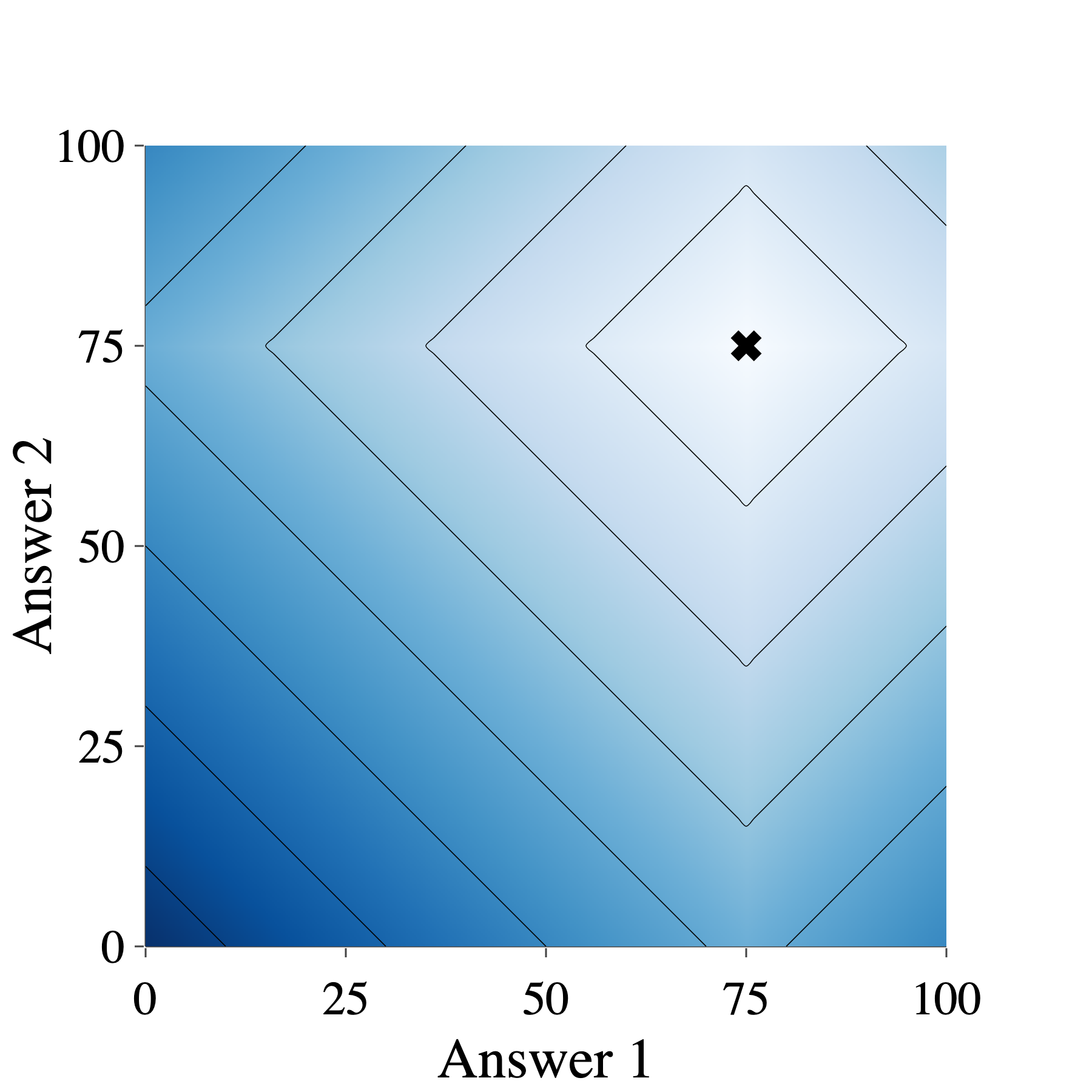}
        \caption{L1}
        \label{fig:L1}
    \end{subfigure}
    \hfill
    \begin{subfigure}{0.24\linewidth}  %
        \centering
        \includegraphics[width=\linewidth]{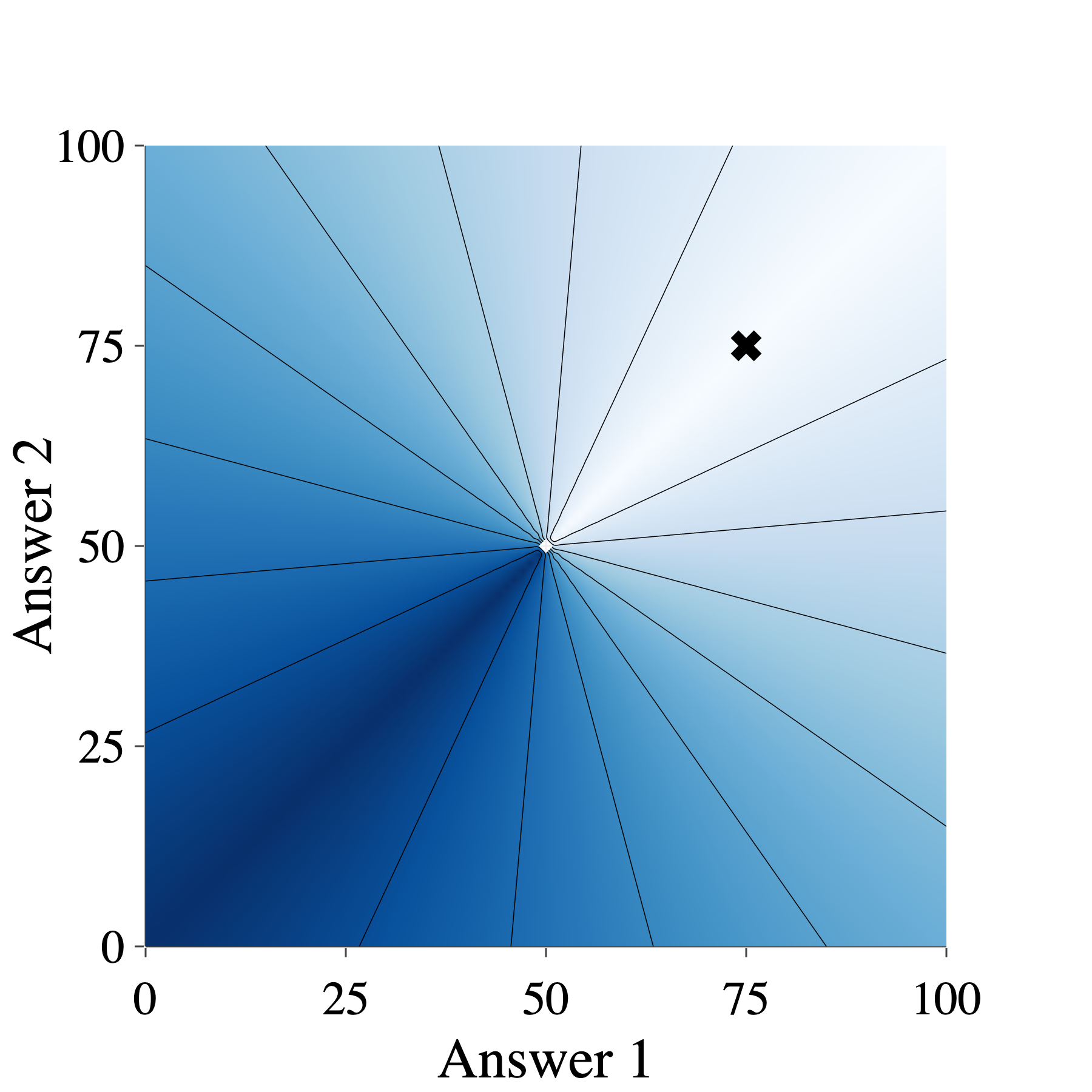}
        \caption{Angular}
        \label{fig:angular}
    \end{subfigure}
    \hfill
    \begin{subfigure}{0.24\linewidth}  %
        \centering
        \includegraphics[width=\linewidth]{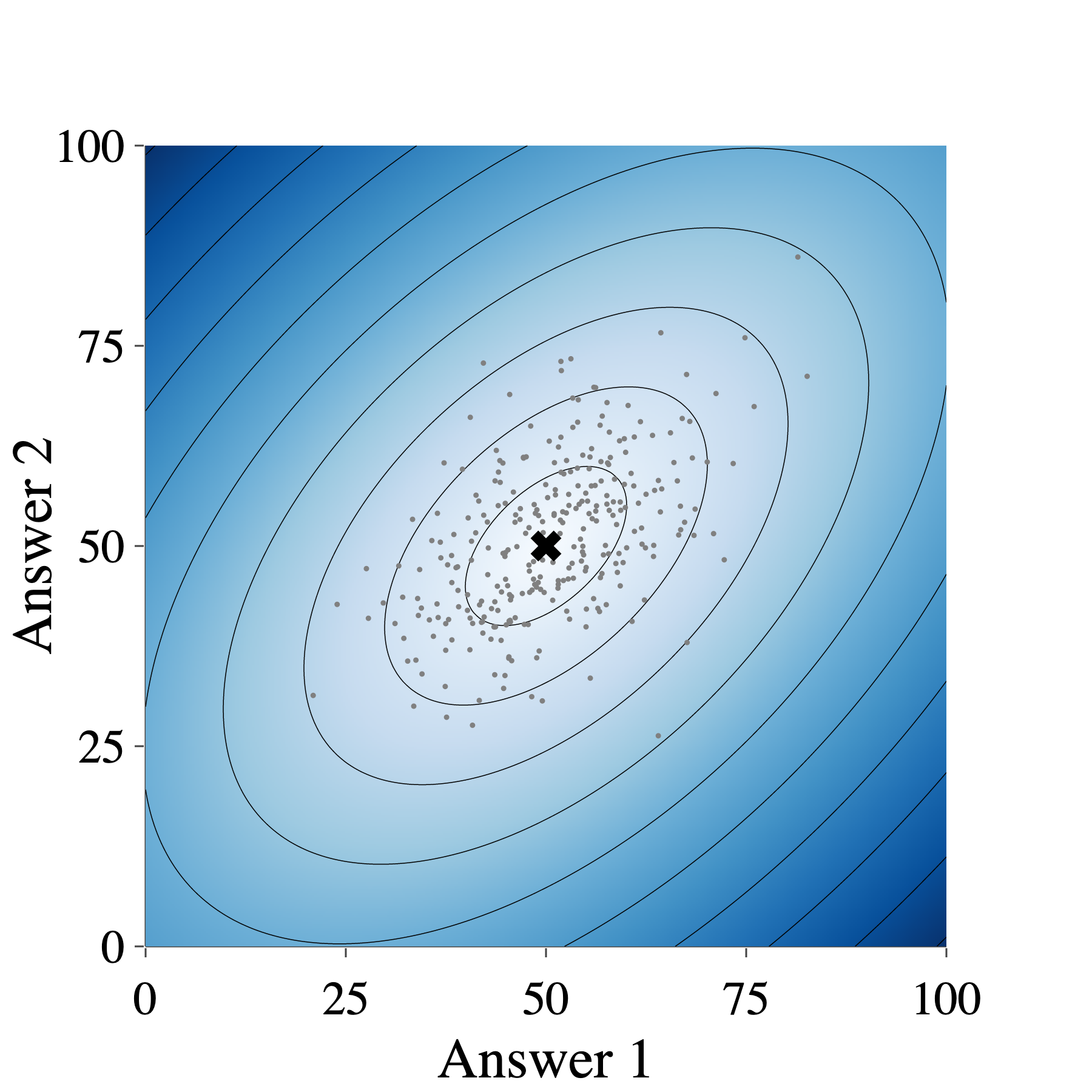}
        \caption{Mahalanobis}
        \label{fig:mahalanobis}
    \end{subfigure}
    \caption{Visualizations of the L2, L1, Angular, and Mahalanobis distance metrics in a simplified answer space with two questions. The contour plot illustrates how the distance of a candidate behaves based on their relative position to the voter, who is represented as \faTimes. For the Mahalanobis distance, the positions of the candidates used for calculating the precision matrix are additionally visualized as small gray dots.}
    \label{fig:distance_metrics}
\end{figure*}

\subsubsection{L1}
The L1 distance, also known as the City-Block or Manhattan distance (see Figure~\ref{fig:L1}), is defined as the sum of the absolute differences across all dimensions, making it a simple and intuitive metric:
\[
d_{\text{L1}}(\mathbf{v}_i, \mathbf{w}_i, \mathbf{c}_j) = \sum_{t=1}^{N_q} \mathbf{w}_{i,t} \cdot |\mathbf{v}_{i,t} - \mathbf{c}_{j,t}|.
\]

\subsubsection{Agreement Count}
The Agreement Count metric simply counts the agreements between voter and candidate responses. To convert it into a distance metric, we take the negative value of the count and incorporate voter weights.
\[
d_{\text{AC}}(\mathbf{v}_i, \mathbf{w}_i, \mathbf{c}_j) = \sum_{t=1}^{N_q} -\mathbf{w}_{i,t} \cdot \delta_{\mathbf{v}_{i,t} \mathbf{c}_{j,t}}
\]
where
\[
\delta_{xy} = 
\begin{cases} 
1 & \text{if } x = y, \\
0 & \text{otherwise}.
\end{cases}
\]
A bias can be added to make the distance positive.
\subsubsection{Angular}
The Angular distance measures the angle between the dimension-weighted voter and candidate answer vectors. This metric captures the directional similarity between the voter’s and candidate’s responses, putting less emphasis on the magnitude of their deviation from the neutral point in comparison to other distance metrics. Assuming that the neutral response vector \(50 \cdot \mathbf{1}_{N_q}\) serves as the origin of the answer space, it is defined as
\[
d_{\text{Angular}}(\mathbf{v}_i, \mathbf{w}_i, \mathbf{c}_j) = \arccos \left( \frac{\tilde{\mathbf{v}}_i^\top \tilde{\mathbf{c}}_j}{\|\tilde{\mathbf{v}}_i\| \cdot \|\tilde{\mathbf{c}}_j\|} \right),
\]
where
\[
\tilde{\mathbf{v}}_i = \mathbf{w}_i \odot (\mathbf{v}_i - 50), \qquad
\tilde{\mathbf{c}}_j = \mathbf{w}_i \odot (\mathbf{c}_j - 50).
\]
Here, \(\odot\) denotes element-wise multiplication. See Figure~\ref{fig:angular} for a visualization.

\subsubsection{Mahalanobis}
The Mahalanobis distance is a covariance-rescaled version of the Euclidean distance. It accounts for the correlations between questions based on the candidate answer vectors and adjusts the importance of each dimension accordingly. Unlike the other distance metrics, the Mahalanobis distance does not take into account the question weights of voters. Figure~\ref{fig:mahalanobis} illustrates how the Mahalanobis distance behaves in a simplified answer space.
\[
d_{\text{Mahalanobis}}(\mathbf{v}_i, \mathbf{c}_j) = \sqrt{(\mathbf{v}_i - \mathbf{c}_j)^\top \text{Cov}(\mathbf{C})^{-1} (\mathbf{v}_i - \mathbf{c}_j)},
\]
where \(\mathbf{C}\) is the candidate answer matrix, with \(\mathbf{C}_j\) representing the \(j\)-th row of \(\mathbf{C}\), corresponding to the answer vector \(\mathbf{c}_j\) of candidate \(j\). The covariance matrix \(\text{Cov}(\mathbf{C})\) is defined as
\[
\text{Cov}(\mathbf{C}) = \frac{1}{N_q - 1} (\mathbf{C} - \bar{\mathbf{C}})^\top (\mathbf{C} - \bar{\mathbf{C}}),
\]
with \(\bar{\mathbf{C}}\) the matrix containing the column means of the candidate answer matrix \(\mathbf{C}\).

\subsubsection{L1 Bonus}
Distance matrices offer an alternative approach to defining distances by specifying the distance between every possible pair of responses in a matrix format. Until 2010, Smartvote utilized a modified version of the L1 distance, known as L1 with Bonus, which employed such a distance matrix for calculating recommendations. This method ensures that the row and column sums of the distance matrix are equal by adding a bonus when voters and candidates strongly agree on a question. The distance for a voter \(i\) and candidate \(j\) under this method is given by:
\[
d_{\text{L1 Bonus}}(\mathbf{v}_i, \mathbf{w}_i, \mathbf{c}_j) = \sum_{t=1}^{N_q} \mathbf{w}_{i,t} \cdot \mathbf{D}_{\text{L1 Bonus}}(\mathbf{v}_{i,t}, \mathbf{c}_{j,t}),
\]
where \(\mathbf{D}_{\text{L1 Bonus}}(\mathbf{v}_{i,t}, \mathbf{c}_{j,t})\) is the entry from the distance matrix \(\mathbf{D}_{\text{L1 Bonus}}\) corresponding to the voter's response \(\mathbf{v}_{i,t}\) and the candidate's response \(\mathbf{c}_{j,t}\) to question \(t\). Table~\ref{tab:l1_bonus_distance_matrix} illustrates \(\mathbf{D}_{\text{L1 Bonus}}\) for questions with five answer options.

\begin{table}[t]
    \centering
    \small
    \begin{tabularx}{\linewidth}{l*{5}{>{\centering\arraybackslash}X}}
        \toprule
        \diagbox{$q(v)$}{$q(c)$} & 0 & 25 & 50 & 75 & 100 \\
        \midrule
        0 & \cellcolor{Blues1}0 & \cellcolor{Blues5}125 & \cellcolor{Blues6}150 & \cellcolor{Blues7}175 & \cellcolor{Blues8}200 \\
        25 & \cellcolor{Blues5}125 & \cellcolor{Blues3}75 & \cellcolor{Blues5}125 & \cellcolor{Blues6}150 & \cellcolor{Blues7}175 \\
        50 & \cellcolor{Blues6}150 & \cellcolor{Blues5}125 & \cellcolor{Blues4}100 & \cellcolor{Blues5}125 & \cellcolor{Blues6}150 \\
        75 & \cellcolor{Blues7}175 & \cellcolor{Blues6}150 & \cellcolor{Blues5}125 & \cellcolor{Blues3}75 & \cellcolor{Blues5}125 \\
        100 & \cellcolor{Blues8}200 & \cellcolor{Blues7}175 & \cellcolor{Blues6}150 & \cellcolor{Blues5}125 & \cellcolor{Blues1}0 \\
        \bottomrule
    \end{tabularx}
    \caption{The L1 Bonus distance matrix used by Smartvote until 2010. The table shows how the distance between a voter and candidate is determined for all pairs of voter $q(v)$ and candidate $q(c)$ responses to a question with five options.}
    \label{tab:l1_bonus_distance_matrix}
\end{table}

\subsubsection{Hybrid}
The Hybrid distance method, used by the EUVox VAA in 2014 \citep{mendez2012hybrid}, balances the proximity voting logic model, which generally focuses on how close candidate and voter answer vectors are, and the directional voting logic model, which emphasizes their similarity in direction over the exact strength of the deviations from the neutral point. The Hybrid distance is calculated as the average of the L1 and scalar distance matrices. The distance for a voter \(i\) and candidate \(j\) under this method is given by
\[
d_{\text{Hybrid}}(\mathbf{v}_i, \mathbf{w}_i, \mathbf{c}_j) = \sum_{t=1}^{N_q} \mathbf{w}_{i,t} \cdot \mathbf{D}_{\text{Hybrid}}(\mathbf{v}_{i,t}, \mathbf{c}_{j,t}),
\]
where \(\mathbf{D}_{\text{Hybrid}}(\mathbf{v}_{i,t}, \mathbf{c}_{j,t})\) is the entry from the Hybrid distance matrix \(\mathbf{D}_{\text{Hybrid}}\) corresponding to the voter's response \(\mathbf{v}_{i,t}\) and the candidate's response \(\mathbf{c}_{j,t}\) to question \(t\). Table~\ref{tab:hybrid_distance_matrix} illustrates \(\mathbf{D}_{\text{Hybrid}}\) for questions with five answer options.

\section{Mitigation Strategies (continued)}
\label{app:mitigations}
\subsubsection{Question Order Randomization (\textcolor{platform}{QO}).}
The question order vulnerability can be effectively mitigated by randomizing the sequence of questions. Importantly, this randomization does not impact the reproducibility property as long as the matching method is permutation invariant. This is generally true for most VAAs, including Smartvote, as observed in Eq.~\eqref{eq:L2_distance} and \eqref{eq:similarity_score}.
\begin{figure}[h]
    \centering
    \includegraphics[width=\linewidth]{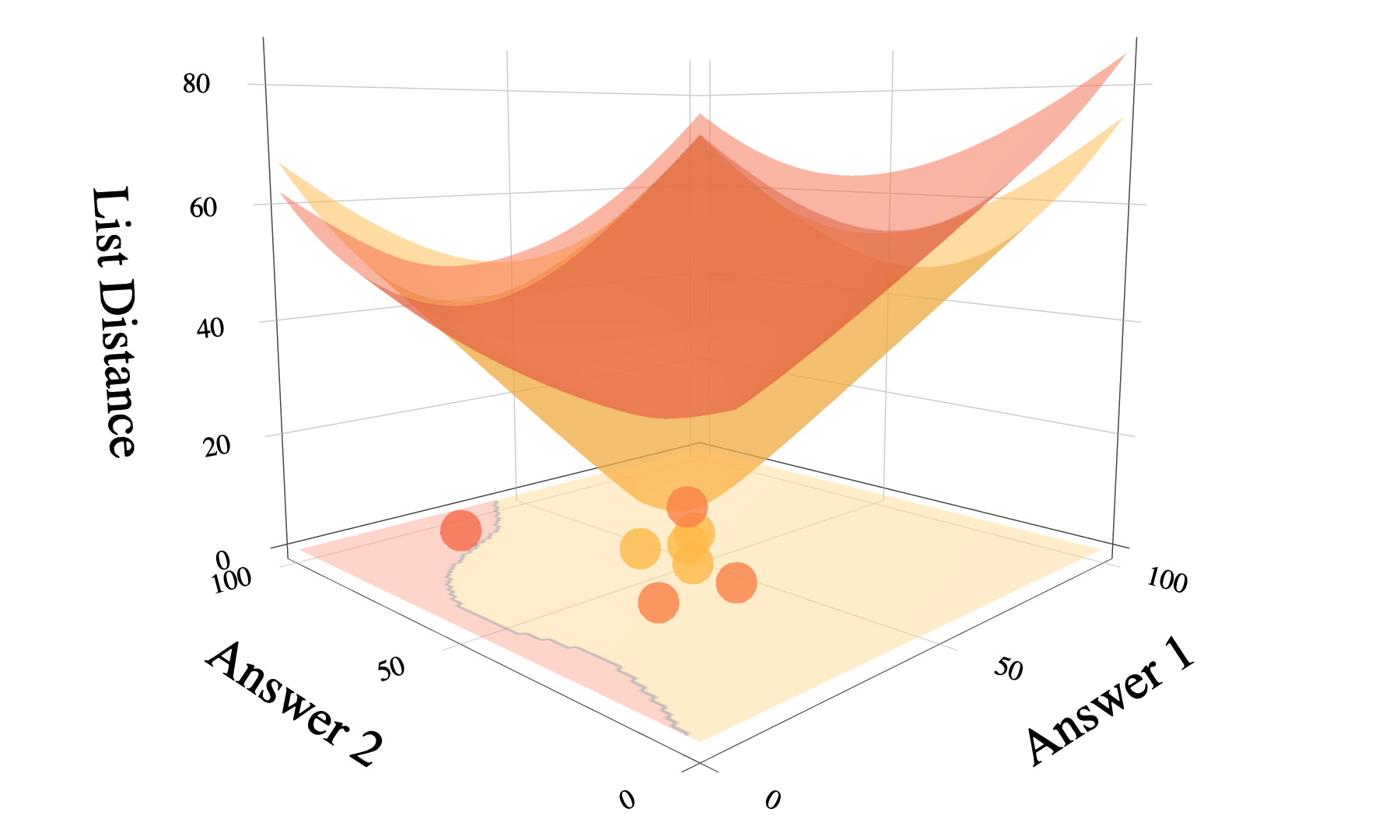}
    \caption{Impact of list centralization on recommendation outcomes. This plot illustrates how a more centralized list, with candidates that have less spread (yellow), is closer to voters in a larger portion of the answer space compared to a list with more spread-out candidates (red).}
    \label{fig:list_centralization}
\end{figure}

\subsubsection{Fair Tie-Breaking (\textcolor{platform}{TB}).}
To ensure fairness in tie-breaking and prevent systematic bias (for instance based on candidates' last names as in Smartvote), we propose generating random seeds for voters based on their ID (to maintain reproducibility). This introduces variability in tie-breaking, ensuring that ties between candidates are not systematically resolved in favor of the same candidate for all voters.

\subsubsection{List Matching Score (\textcolor{candidate}{LC}).}
Smartvote ranks lists by computing the average similarity scores between the voter and the list candidates. Figure~\ref{fig:list_centralization} exposes the mechanism driving the list centralization bias: When candidates on a list have similar answer profiles, the average distance to the voter is minimized, increasing the likelihood of the list being recommended. In contrast, lists with diverse candidate responses incur a higher average distance, reducing their chances of recommendation. 
Instead of ranking lists by the average similarity scores between the voter and the list candidates, we propose ranking them by the similarity between the voter and the average answer vector of the candidates on the list. This approach would effectively neutralize the impact of list centralization strategies.

\begin{table}[t]
    \centering
    \small
    \begin{tabularx}{\linewidth}{l*{5}{>{\centering\arraybackslash}X}}
        \toprule
        \diagbox{$q(v)$}{$q(c)$} & 0 & 25 & 50 & 75 & 100 \\
        \midrule
        0 & \cellcolor{Blues1}0 & \cellcolor{Blues3}50 & \cellcolor{Blues5}100 & \cellcolor{Blues7}150 & \cellcolor{Blues8}200 \\
        25 & \cellcolor{Blues3}50 & \cellcolor{Blues2}37.5 & \cellcolor{Blues4}75 & \cellcolor{Blues6}112.5 & \cellcolor{Blues7}150 \\
        50 & \cellcolor{Blues5}100 & \cellcolor{Blues4}75 & \cellcolor{Blues3}50 & \cellcolor{Blues4}75 & \cellcolor{Blues5}100 \\
        75 & \cellcolor{Blues7}150 & \cellcolor{Blues6}112.5 & \cellcolor{Blues4}75 & \cellcolor{Blues2}37.5 & \cellcolor{Blues3}50 \\
        100 & \cellcolor{Blues8}200 & \cellcolor{Blues7}150 & \cellcolor{Blues5}100 & \cellcolor{Blues3}50 & \cellcolor{Blues1}0 \\
        \bottomrule
    \end{tabularx}
    \caption{The Hybrid distance matrix used by the EUVox 2014 VAA. The table shows how the distance between a voter and candidate is determined for all pairs of voter $q(v)$ and candidate $q(c)$ responses to a question with five options.}
    \label{tab:hybrid_distance_matrix}
\end{table}

\begin{figure}[h]
    \centering
    \includegraphics[width=1\linewidth]{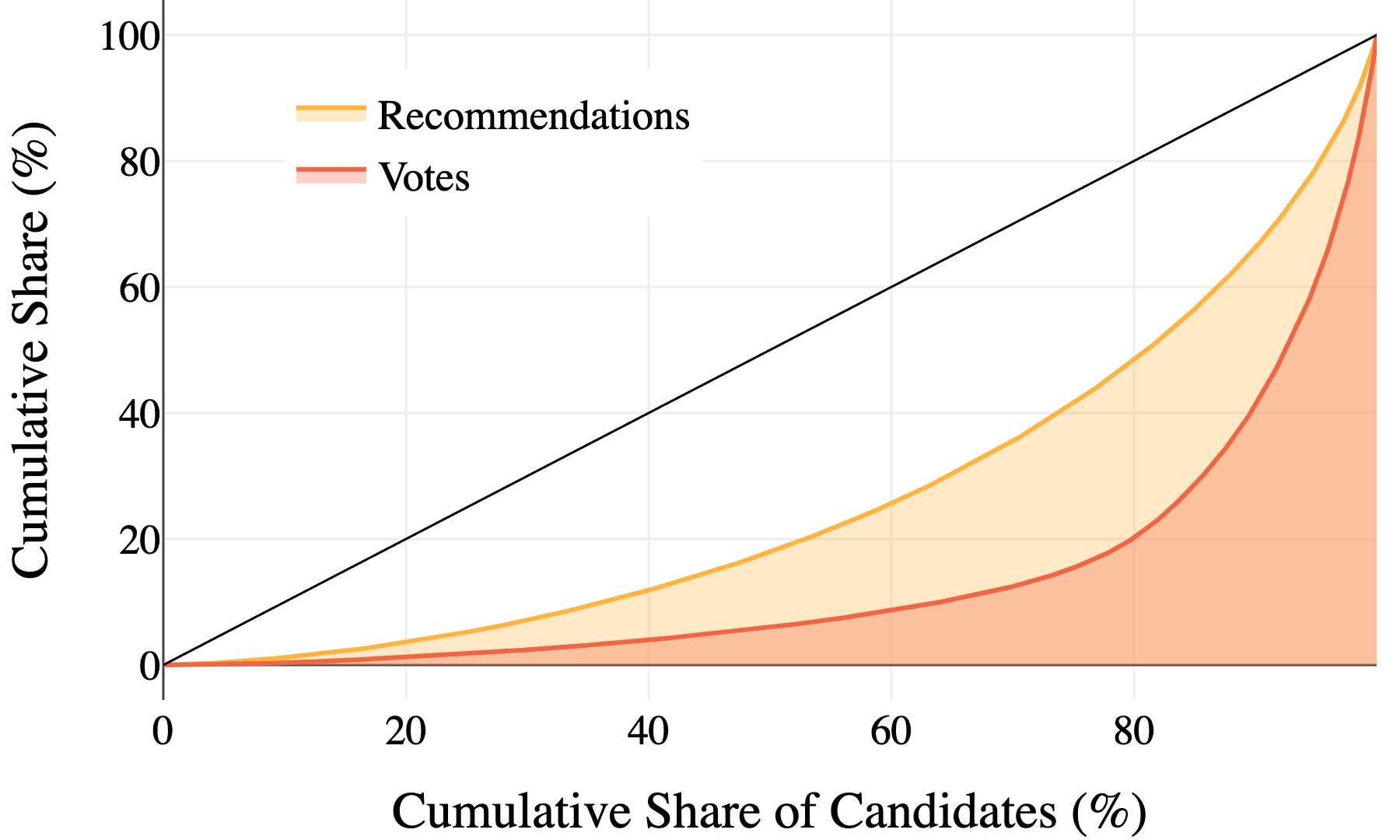}
    \caption{Cumulative share of expectation-normalized visibility and expectation-normalized actual votes over all candidates. The Lorenz curves illustrate that actual votes are distributed less evenly than Smartvote recommendations over all candidates.}
    \label{fig:gini}
\end{figure}

\section{Related Work (continued)}
\label{app:related_work}

\subsubsection{VAA Impact on Democracy.}
Post-electoral surveys analyzed by \citet{ladner2012voting} indicate that 67\% of voters are influenced by Smartvote, with 15\% of voters stating they had adopted the recommendation in its entirety. Moreover, the candidate recommendations by Smartvote have been shown to be the most decisive factor of electoral choice, ahead of e.g., party membership~\citep{ladner2016promissory}.
\citet{alvarez2014useful} found that 4 out of 5 users of the \textit{EU Profiler} evaluated the VAA as useful. Interestingly, perceived usefulness decreased whenever a user's preferences were less represented. For the same dataset, \citet{alvarez2014impact} find that more than 4 in 5 voters were matched with a party they did not initially prefer, with 8\% of those voters subsequently changing their vote.
\citet{germann2019getting} estimate that in 2007 a staggering 58'000 voters were added through the existence of smartvote, making up about 1.2\% of the total tally. They compute the cost of each additional voter to be 7.5 USD, exceptionally low compared to costs averaging between 38 and 90 USD in telephone campaigns.
Finally, \citet{munzert2021meta} perform a meta-analysis of the effects of VAAs and show that there is significant evidence that VAAs increase voter turnout and influence vote choice. 

\subsubsection{VAA Development.}
\citet{garziamarschall2014matching} collect an extensive overview of the state-of-the-art on VAA research. 
\citet{mendez2017modeling} compares the predictive power of both high- and low-dimensional matching methods, by using the voter's initial preferences as ground truth. 
\citet{garzia2019open} postulates a set of open questions in VAA research, underlining that identifying which matching algorithms are best suited is still an open question. Our study does not give all answers but provides strong evidence that certain matching algorithms should not be used.

\end{document}